\documentclass[longauth]{aa}
\usepackage{graphicx}
\usepackage{natbib}
\usepackage{scalerel}
\usepackage{lastpage}

\usepackage{txfonts}
\usepackage{hyperref}
\hypersetup{
    colorlinks=true,
    linkcolor=blue,
    filecolor=magenta,      
    urlcolor=blue,
    citecolor=blue
}
\makeatletter
\renewcommand*\aa@pageof{, page \thepage{} of \pageref*{LastPage}}
\makeatother
\usepackage[utf8]{inputenc}
\IfFileExists{orcidlink.sty}{
  \usepackage{orcidlink}
  \newcommand{\orcid}[1]{\orcidlink{##1}}
}{
  \newcommand{\orcid}[1]{}
}
\usepackage{euclid}
\usepackage{multirow}
\usepackage{lineno}

\begin{document}

\title{\Euclid preparation}
\subtitle{Decomposing components of the extragalactic background light using multi-band intensity mapping cross-correlations}    
\author{Euclid Collaboration: Y.~Cao\orcid{0000-0001-6111-4157}\thanks{\email{yec15@uci.edu}}\inst{\ref{aff1}}
\and A.~R.~Cooray\orcid{0000-0002-3892-0190}\inst{\ref{aff1}}
\and T.~Li\orcid{0009-0003-5064-1914}\inst{\ref{aff1}}
\and Y.-T.~Cheng\orcid{0000-0002-5437-0504}\inst{\ref{aff2}}
\and K.~Tanidis\orcid{0000-0001-9843-5130}\inst{\ref{aff3}}
\and S.~H.~Lim\orcid{0000-0001-6860-9064}\inst{\ref{aff4},\ref{aff5}}
\and D.~Scott\orcid{0000-0002-6878-9840}\inst{\ref{aff6}}
\and B.~Altieri\orcid{0000-0003-3936-0284}\inst{\ref{aff7}}
\and A.~Amara\inst{\ref{aff8}}
\and S.~Andreon\orcid{0000-0002-2041-8784}\inst{\ref{aff9}}
\and N.~Auricchio\orcid{0000-0003-4444-8651}\inst{\ref{aff10}}
\and C.~Baccigalupi\orcid{0000-0002-8211-1630}\inst{\ref{aff11},\ref{aff12},\ref{aff13},\ref{aff14}}
\and M.~Baldi\orcid{0000-0003-4145-1943}\inst{\ref{aff15},\ref{aff10},\ref{aff16}}
\and S.~Bardelli\orcid{0000-0002-8900-0298}\inst{\ref{aff10}}
\and A.~Biviano\orcid{0000-0002-0857-0732}\inst{\ref{aff12},\ref{aff11}}
\and E.~Branchini\orcid{0000-0002-0808-6908}\inst{\ref{aff17},\ref{aff18},\ref{aff9}}
\and M.~Brescia\orcid{0000-0001-9506-5680}\inst{\ref{aff19},\ref{aff20}}
\and S.~Camera\orcid{0000-0003-3399-3574}\inst{\ref{aff21},\ref{aff22},\ref{aff23}}
\and G.~Ca\~nas-Herrera\orcid{0000-0003-2796-2149}\inst{\ref{aff24},\ref{aff25}}
\and V.~Capobianco\orcid{0000-0002-3309-7692}\inst{\ref{aff23}}
\and C.~Carbone\orcid{0000-0003-0125-3563}\inst{\ref{aff26}}
\and J.~Carretero\orcid{0000-0002-3130-0204}\inst{\ref{aff27},\ref{aff28}}
\and S.~Casas\orcid{0000-0002-4751-5138}\inst{\ref{aff29},\ref{aff30}}
\and M.~Castellano\orcid{0000-0001-9875-8263}\inst{\ref{aff31}}
\and G.~Castignani\orcid{0000-0001-6831-0687}\inst{\ref{aff10}}
\and S.~Cavuoti\orcid{0000-0002-3787-4196}\inst{\ref{aff20},\ref{aff32}}
\and K.~C.~Chambers\orcid{0000-0001-6965-7789}\inst{\ref{aff33}}
\and A.~Cimatti\inst{\ref{aff34}}
\and C.~Colodro-Conde\inst{\ref{aff35}}
\and G.~Congedo\orcid{0000-0003-2508-0046}\inst{\ref{aff24}}
\and C.~J.~Conselice\orcid{0000-0003-1949-7638}\inst{\ref{aff36}}
\and L.~Conversi\orcid{0000-0002-6710-8476}\inst{\ref{aff37},\ref{aff7}}
\and Y.~Copin\orcid{0000-0002-5317-7518}\inst{\ref{aff38}}
\and F.~Courbin\orcid{0000-0003-0758-6510}\inst{\ref{aff39},\ref{aff40},\ref{aff41}}
\and H.~M.~Courtois\orcid{0000-0003-0509-1776}\inst{\ref{aff42}}
\and J.-C.~Cuillandre\orcid{0000-0002-3263-8645}\inst{\ref{aff43}}
\and H.~Degaudenzi\orcid{0000-0002-5887-6799}\inst{\ref{aff44}}
\and G.~De~Lucia\orcid{0000-0002-6220-9104}\inst{\ref{aff12}}
\and H.~Dole\orcid{0000-0002-9767-3839}\inst{\ref{aff45}}
\and M.~Douspis\orcid{0000-0003-4203-3954}\inst{\ref{aff45}}
\and F.~Dubath\orcid{0000-0002-6533-2810}\inst{\ref{aff44}}
\and X.~Dupac\inst{\ref{aff7}}
\and M.~Farina\orcid{0000-0002-3089-7846}\inst{\ref{aff46}}
\and R.~Farinelli\inst{\ref{aff10}}
\and F.~Faustini\orcid{0000-0001-6274-5145}\inst{\ref{aff31},\ref{aff47}}
\and S.~Ferriol\inst{\ref{aff38}}
\and F.~Finelli\orcid{0000-0002-6694-3269}\inst{\ref{aff10},\ref{aff48}}
\and M.~Frailis\orcid{0000-0002-7400-2135}\inst{\ref{aff12}}
\and E.~Franceschi\orcid{0000-0002-0585-6591}\inst{\ref{aff10}}
\and M.~Fumana\orcid{0000-0001-6787-5950}\inst{\ref{aff26}}
\and S.~Galeotta\orcid{0000-0002-3748-5115}\inst{\ref{aff12}}
\and K.~George\orcid{0000-0002-1734-8455}\inst{\ref{aff49}}
\and B.~Gillis\orcid{0000-0002-4478-1270}\inst{\ref{aff24}}
\and C.~Giocoli\orcid{0000-0002-9590-7961}\inst{\ref{aff10},\ref{aff16}}
\and J.~Gracia-Carpio\inst{\ref{aff50}}
\and A.~Grazian\orcid{0000-0002-5688-0663}\inst{\ref{aff51}}
\and F.~Grupp\inst{\ref{aff50},\ref{aff52}}
\and S.~V.~H.~Haugan\orcid{0000-0001-9648-7260}\inst{\ref{aff53}}
\and W.~Holmes\inst{\ref{aff54}}
\and F.~Hormuth\inst{\ref{aff55}}
\and A.~Hornstrup\orcid{0000-0002-3363-0936}\inst{\ref{aff56},\ref{aff57}}
\and K.~Jahnke\orcid{0000-0003-3804-2137}\inst{\ref{aff58}}
\and M.~Jhabvala\inst{\ref{aff59}}
\and B.~Joachimi\orcid{0000-0001-7494-1303}\inst{\ref{aff60}}
\and S.~Kermiche\orcid{0000-0002-0302-5735}\inst{\ref{aff61}}
\and A.~Kiessling\orcid{0000-0002-2590-1273}\inst{\ref{aff54}}
\and B.~Kubik\orcid{0009-0006-5823-4880}\inst{\ref{aff38}}
\and M.~Kunz\orcid{0000-0002-3052-7394}\inst{\ref{aff62}}
\and H.~Kurki-Suonio\orcid{0000-0002-4618-3063}\inst{\ref{aff63},\ref{aff64}}
\and A.~M.~C.~Le~Brun\orcid{0000-0002-0936-4594}\inst{\ref{aff65}}
\and S.~Ligori\orcid{0000-0003-4172-4606}\inst{\ref{aff23}}
\and P.~B.~Lilje\orcid{0000-0003-4324-7794}\inst{\ref{aff53}}
\and V.~Lindholm\orcid{0000-0003-2317-5471}\inst{\ref{aff63},\ref{aff64}}
\and I.~Lloro\orcid{0000-0001-5966-1434}\inst{\ref{aff66}}
\and G.~Mainetti\orcid{0000-0003-2384-2377}\inst{\ref{aff67}}
\and D.~Maino\inst{\ref{aff68},\ref{aff26},\ref{aff69}}
\and E.~Maiorano\orcid{0000-0003-2593-4355}\inst{\ref{aff10}}
\and O.~Mansutti\orcid{0000-0001-5758-4658}\inst{\ref{aff12}}
\and S.~Marcin\inst{\ref{aff70}}
\and O.~Marggraf\orcid{0000-0001-7242-3852}\inst{\ref{aff71}}
\and M.~Martinelli\orcid{0000-0002-6943-7732}\inst{\ref{aff31},\ref{aff72}}
\and N.~Martinet\orcid{0000-0003-2786-7790}\inst{\ref{aff73}}
\and F.~Marulli\orcid{0000-0002-8850-0303}\inst{\ref{aff74},\ref{aff10},\ref{aff16}}
\and R.~J.~Massey\orcid{0000-0002-6085-3780}\inst{\ref{aff75}}
\and E.~Medinaceli\orcid{0000-0002-4040-7783}\inst{\ref{aff10}}
\and S.~Mei\orcid{0000-0002-2849-559X}\inst{\ref{aff76},\ref{aff77}}
\and Y.~Mellier\thanks{Deceased}\inst{\ref{aff78},\ref{aff79}}
\and M.~Meneghetti\orcid{0000-0003-1225-7084}\inst{\ref{aff10},\ref{aff16}}
\and E.~Merlin\orcid{0000-0001-6870-8900}\inst{\ref{aff31}}
\and G.~Meylan\inst{\ref{aff80}}
\and A.~Mora\orcid{0000-0002-1922-8529}\inst{\ref{aff81}}
\and M.~Moresco\orcid{0000-0002-7616-7136}\inst{\ref{aff74},\ref{aff10}}
\and L.~Moscardini\orcid{0000-0002-3473-6716}\inst{\ref{aff74},\ref{aff10},\ref{aff16}}
\and C.~Neissner\orcid{0000-0001-8524-4968}\inst{\ref{aff82},\ref{aff28}}
\and S.-M.~Niemi\orcid{0009-0005-0247-0086}\inst{\ref{aff83}}
\and C.~Padilla\orcid{0000-0001-7951-0166}\inst{\ref{aff82}}
\and S.~Paltani\orcid{0000-0002-8108-9179}\inst{\ref{aff44}}
\and F.~Pasian\orcid{0000-0002-4869-3227}\inst{\ref{aff12}}
\and K.~Pedersen\inst{\ref{aff84}}
\and W.~J.~Percival\orcid{0000-0002-0644-5727}\inst{\ref{aff85},\ref{aff86},\ref{aff87}}
\and V.~Pettorino\orcid{0000-0002-4203-9320}\inst{\ref{aff83}}
\and G.~Polenta\orcid{0000-0003-4067-9196}\inst{\ref{aff47}}
\and M.~Poncet\inst{\ref{aff88}}
\and L.~A.~Popa\inst{\ref{aff89}}
\and F.~Raison\orcid{0000-0002-7819-6918}\inst{\ref{aff50}}
\and A.~Renzi\orcid{0000-0001-9856-1970}\inst{\ref{aff90},\ref{aff91}}
\and J.~Rhodes\orcid{0000-0002-4485-8549}\inst{\ref{aff54}}
\and G.~Riccio\inst{\ref{aff20}}
\and E.~Romelli\orcid{0000-0003-3069-9222}\inst{\ref{aff12}}
\and M.~Roncarelli\orcid{0000-0001-9587-7822}\inst{\ref{aff10}}
\and R.~Saglia\orcid{0000-0003-0378-7032}\inst{\ref{aff52},\ref{aff50}}
\and Z.~Sakr\orcid{0000-0002-4823-3757}\inst{\ref{aff92},\ref{aff93},\ref{aff94}}
\and D.~Sapone\orcid{0000-0001-7089-4503}\inst{\ref{aff95}}
\and P.~Schneider\orcid{0000-0001-8561-2679}\inst{\ref{aff71}}
\and T.~Schrabback\orcid{0000-0002-6987-7834}\inst{\ref{aff96}}
\and A.~Secroun\orcid{0000-0003-0505-3710}\inst{\ref{aff61}}
\and G.~Seidel\orcid{0000-0003-2907-353X}\inst{\ref{aff58}}
\and S.~Serrano\orcid{0000-0002-0211-2861}\inst{\ref{aff97},\ref{aff98},\ref{aff99}}
\and E.~Sihvola\orcid{0000-0003-1804-7715}\inst{\ref{aff100}}
\and C.~Sirignano\orcid{0000-0002-0995-7146}\inst{\ref{aff90},\ref{aff91}}
\and G.~Sirri\orcid{0000-0003-2626-2853}\inst{\ref{aff16}}
\and L.~Stanco\orcid{0000-0002-9706-5104}\inst{\ref{aff91}}
\and J.~Steinwagner\orcid{0000-0001-7443-1047}\inst{\ref{aff50}}
\and P.~Tallada-Cresp\'{i}\orcid{0000-0002-1336-8328}\inst{\ref{aff27},\ref{aff28}}
\and I.~Tereno\orcid{0000-0002-4537-6218}\inst{\ref{aff101},\ref{aff102}}
\and N.~Tessore\orcid{0000-0002-9696-7931}\inst{\ref{aff103}}
\and S.~Toft\orcid{0000-0003-3631-7176}\inst{\ref{aff104},\ref{aff105}}
\and R.~Toledo-Moreo\orcid{0000-0002-2997-4859}\inst{\ref{aff106}}
\and F.~Torradeflot\orcid{0000-0003-1160-1517}\inst{\ref{aff28},\ref{aff27}}
\and I.~Tutusaus\orcid{0000-0002-3199-0399}\inst{\ref{aff99},\ref{aff97},\ref{aff93}}
\and L.~Valenziano\orcid{0000-0002-1170-0104}\inst{\ref{aff10},\ref{aff48}}
\and J.~Valiviita\orcid{0000-0001-6225-3693}\inst{\ref{aff63},\ref{aff64}}
\and T.~Vassallo\orcid{0000-0001-6512-6358}\inst{\ref{aff12}}
\and Y.~Wang\orcid{0000-0002-4749-2984}\inst{\ref{aff107}}
\and J.~Weller\orcid{0000-0002-8282-2010}\inst{\ref{aff52},\ref{aff50}}
\and G.~Zamorani\orcid{0000-0002-2318-301X}\inst{\ref{aff10}}
\and F.~M.~Zerbi\inst{\ref{aff9}}
\and E.~Zucca\orcid{0000-0002-5845-8132}\inst{\ref{aff10}}
\and M.~Ballardini\orcid{0000-0003-4481-3559}\inst{\ref{aff108},\ref{aff109},\ref{aff10}}
\and E.~Bozzo\orcid{0000-0002-8201-1525}\inst{\ref{aff44}}
\and C.~Burigana\orcid{0000-0002-3005-5796}\inst{\ref{aff110},\ref{aff48}}
\and R.~Cabanac\orcid{0000-0001-6679-2600}\inst{\ref{aff93}}
\and M.~Calabrese\orcid{0000-0002-2637-2422}\inst{\ref{aff111},\ref{aff26}}
\and A.~Cappi\inst{\ref{aff112},\ref{aff10}}
\and T.~Castro\orcid{0000-0002-6292-3228}\inst{\ref{aff12},\ref{aff13},\ref{aff11},\ref{aff113}}
\and J.~A.~Escartin~Vigo\inst{\ref{aff50}}
\and L.~Gabarra\orcid{0000-0002-8486-8856}\inst{\ref{aff114}}
\and J.~Macias-Perez\orcid{0000-0002-5385-2763}\inst{\ref{aff115}}
\and R.~Maoli\orcid{0000-0002-6065-3025}\inst{\ref{aff116},\ref{aff31}}
\and J.~Mart\'{i}n-Fleitas\orcid{0000-0002-8594-569X}\inst{\ref{aff117}}
\and N.~Mauri\orcid{0000-0001-8196-1548}\inst{\ref{aff34},\ref{aff16}}
\and R.~B.~Metcalf\orcid{0000-0003-3167-2574}\inst{\ref{aff74},\ref{aff10}}
\and P.~Monaco\orcid{0000-0003-2083-7564}\inst{\ref{aff118},\ref{aff12},\ref{aff13},\ref{aff11}}
\and A.~A.~Nucita\inst{\ref{aff119},\ref{aff120},\ref{aff121}}
\and A.~Pezzotta\orcid{0000-0003-0726-2268}\inst{\ref{aff9}}
\and M.~P\"ontinen\orcid{0000-0001-5442-2530}\inst{\ref{aff63}}
\and I.~Risso\orcid{0000-0003-2525-7761}\inst{\ref{aff9},\ref{aff18}}
\and V.~Scottez\orcid{0009-0008-3864-940X}\inst{\ref{aff78},\ref{aff122}}
\and M.~Sereno\orcid{0000-0003-0302-0325}\inst{\ref{aff10},\ref{aff16}}
\and M.~Tenti\orcid{0000-0002-4254-5901}\inst{\ref{aff16}}
\and M.~Tucci\inst{\ref{aff44}}
\and M.~Viel\orcid{0000-0002-2642-5707}\inst{\ref{aff11},\ref{aff12},\ref{aff14},\ref{aff13},\ref{aff113}}
\and M.~Wiesmann\orcid{0009-0000-8199-5860}\inst{\ref{aff53}}
\and Y.~Akrami\orcid{0000-0002-2407-7956}\inst{\ref{aff123},\ref{aff124}}
\and I.~T.~Andika\orcid{0000-0001-6102-9526}\inst{\ref{aff49}}
\and G.~Angora\orcid{0000-0002-0316-6562}\inst{\ref{aff20},\ref{aff108}}
\and S.~Anselmi\orcid{0000-0002-3579-9583}\inst{\ref{aff91},\ref{aff90},\ref{aff125}}
\and M.~Archidiacono\orcid{0000-0003-4952-9012}\inst{\ref{aff68},\ref{aff69}}
\and E.~Aubourg\orcid{0000-0002-5592-023X}\inst{\ref{aff76},\ref{aff126}}
\and L.~Bazzanini\orcid{0000-0003-0727-0137}\inst{\ref{aff108},\ref{aff10}}
\and D.~Bertacca\orcid{0000-0002-2490-7139}\inst{\ref{aff90},\ref{aff51},\ref{aff91}}
\and M.~Bethermin\orcid{0000-0002-3915-2015}\inst{\ref{aff127}}
\and A.~Blanchard\orcid{0000-0001-8555-9003}\inst{\ref{aff93}}
\and L.~Blot\orcid{0000-0002-9622-7167}\inst{\ref{aff128},\ref{aff65}}
\and M.~Bonici\orcid{0000-0002-8430-126X}\inst{\ref{aff85},\ref{aff26}}
\and S.~Borgani\orcid{0000-0001-6151-6439}\inst{\ref{aff118},\ref{aff11},\ref{aff12},\ref{aff13},\ref{aff113}}
\and M.~L.~Brown\orcid{0000-0002-0370-8077}\inst{\ref{aff36}}
\and S.~Bruton\orcid{0000-0002-6503-5218}\inst{\ref{aff2}}
\and A.~Calabro\orcid{0000-0003-2536-1614}\inst{\ref{aff31}}
\and B.~Camacho~Quevedo\orcid{0000-0002-8789-4232}\inst{\ref{aff11},\ref{aff14},\ref{aff12}}
\and F.~Caro\inst{\ref{aff31}}
\and C.~S.~Carvalho\inst{\ref{aff102}}
\and F.~Cogato\orcid{0000-0003-4632-6113}\inst{\ref{aff74},\ref{aff10}}
\and S.~Conseil\orcid{0000-0002-3657-4191}\inst{\ref{aff38}}
\and O.~Cucciati\orcid{0000-0002-9336-7551}\inst{\ref{aff10}}
\and S.~Davini\orcid{0000-0003-3269-1718}\inst{\ref{aff18}}
\and G.~Desprez\orcid{0000-0001-8325-1742}\inst{\ref{aff129}}
\and A.~D\'iaz-S\'anchez\orcid{0000-0003-0748-4768}\inst{\ref{aff130}}
\and S.~Di~Domizio\orcid{0000-0003-2863-5895}\inst{\ref{aff17},\ref{aff18}}
\and J.~M.~Diego\orcid{0000-0001-9065-3926}\inst{\ref{aff131}}
\and V.~Duret\orcid{0009-0009-0383-4960}\inst{\ref{aff61}}
\and M.~Y.~Elkhashab\orcid{0000-0001-9306-2603}\inst{\ref{aff12},\ref{aff13},\ref{aff118},\ref{aff11}}
\and A.~Enia\orcid{0000-0002-0200-2857}\inst{\ref{aff10}}
\and A.~Finoguenov\orcid{0000-0002-4606-5403}\inst{\ref{aff63}}
\and A.~Fontana\orcid{0000-0003-3820-2823}\inst{\ref{aff31}}
\and A.~Franco\orcid{0000-0002-4761-366X}\inst{\ref{aff120},\ref{aff119},\ref{aff121}}
\and K.~Ganga\orcid{0000-0001-8159-8208}\inst{\ref{aff76}}
\and T.~Gasparetto\orcid{0000-0002-7913-4866}\inst{\ref{aff31}}
\and E.~Gaztanaga\orcid{0000-0001-9632-0815}\inst{\ref{aff99},\ref{aff97},\ref{aff132}}
\and F.~Giacomini\orcid{0000-0002-3129-2814}\inst{\ref{aff16}}
\and F.~Gianotti\orcid{0000-0003-4666-119X}\inst{\ref{aff10}}
\and G.~Gozaliasl\orcid{0000-0002-0236-919X}\inst{\ref{aff133},\ref{aff63}}
\and A.~Gruppuso\orcid{0000-0001-9272-5292}\inst{\ref{aff10},\ref{aff16}}
\and M.~Guidi\orcid{0000-0001-9408-1101}\inst{\ref{aff15},\ref{aff10}}
\and C.~M.~Gutierrez\orcid{0000-0001-7854-783X}\inst{\ref{aff35},\ref{aff134}}
\and A.~Hall\orcid{0000-0002-3139-8651}\inst{\ref{aff24}}
\and C.~Hern\'andez-Monteagudo\orcid{0000-0001-5471-9166}\inst{\ref{aff134},\ref{aff35}}
\and H.~Hildebrandt\orcid{0000-0002-9814-3338}\inst{\ref{aff135}}
\and J.~Hjorth\orcid{0000-0002-4571-2306}\inst{\ref{aff84}}
\and J.~J.~E.~Kajava\orcid{0000-0002-3010-8333}\inst{\ref{aff136},\ref{aff137}}
\and Y.~Kang\orcid{0009-0000-8588-7250}\inst{\ref{aff44}}
\and V.~Kansal\orcid{0000-0002-4008-6078}\inst{\ref{aff138},\ref{aff139}}
\and D.~Karagiannis\orcid{0000-0002-4927-0816}\inst{\ref{aff108},\ref{aff140}}
\and K.~Kiiveri\inst{\ref{aff100}}
\and J.~Kim\orcid{0000-0003-2776-2761}\inst{\ref{aff114}}
\and C.~C.~Kirkpatrick\inst{\ref{aff100}}
\and S.~Kruk\orcid{0000-0001-8010-8879}\inst{\ref{aff7}}
\and M.~Lattanzi\orcid{0000-0003-1059-2532}\inst{\ref{aff109}}
\and L.~Legrand\orcid{0000-0003-0610-5252}\inst{\ref{aff141},\ref{aff4}}
\and F.~Lepori\orcid{0009-0000-5061-7138}\inst{\ref{aff142}}
\and G.~Leroy\orcid{0009-0004-2523-4425}\inst{\ref{aff143},\ref{aff75}}
\and G.~F.~Lesci\orcid{0000-0002-4607-2830}\inst{\ref{aff74},\ref{aff10}}
\and J.~Lesgourgues\orcid{0000-0001-7627-353X}\inst{\ref{aff29}}
\and T.~I.~Liaudat\orcid{0000-0002-9104-314X}\inst{\ref{aff126}}
\and S.~J.~Liu\orcid{0000-0001-7680-2139}\inst{\ref{aff46}}
\and M.~Magliocchetti\orcid{0000-0001-9158-4838}\inst{\ref{aff46}}
\and F.~Mannucci\orcid{0000-0002-4803-2381}\inst{\ref{aff144}}
\and C.~J.~A.~P.~Martins\orcid{0000-0002-4886-9261}\inst{\ref{aff145},\ref{aff146}}
\and L.~Maurin\orcid{0000-0002-8406-0857}\inst{\ref{aff45}}
\and M.~Miluzio\inst{\ref{aff7},\ref{aff147}}
\and C.~Moretti\orcid{0000-0003-3314-8936}\inst{\ref{aff12},\ref{aff11},\ref{aff13}}
\and G.~Morgante\inst{\ref{aff10}}
\and K.~Naidoo\orcid{0000-0002-9182-1802}\inst{\ref{aff132},\ref{aff58}}
\and P.~Natoli\orcid{0000-0003-0126-9100}\inst{\ref{aff108},\ref{aff109}}
\and A.~Navarro-Alsina\orcid{0000-0002-3173-2592}\inst{\ref{aff71}}
\and S.~Nesseris\orcid{0000-0002-0567-0324}\inst{\ref{aff123}}
\and L.~Pagano\orcid{0000-0003-1820-5998}\inst{\ref{aff108},\ref{aff109}}
\and D.~Paoletti\orcid{0000-0003-4761-6147}\inst{\ref{aff10},\ref{aff48}}
\and F.~Passalacqua\orcid{0000-0002-8606-4093}\inst{\ref{aff90},\ref{aff91}}
\and K.~Paterson\orcid{0000-0001-8340-3486}\inst{\ref{aff58}}
\and L.~Patrizii\inst{\ref{aff16}}
\and A.~Pisani\orcid{0000-0002-6146-4437}\inst{\ref{aff61}}
\and D.~Potter\orcid{0000-0002-0757-5195}\inst{\ref{aff142}}
\and G.~W.~Pratt\inst{\ref{aff43}}
\and S.~Quai\orcid{0000-0002-0449-8163}\inst{\ref{aff74},\ref{aff10}}
\and M.~Radovich\orcid{0000-0002-3585-866X}\inst{\ref{aff51}}
\and G.~Rodighiero\orcid{0000-0002-9415-2296}\inst{\ref{aff90},\ref{aff51}}
\and K.~Rojas\orcid{0000-0003-1391-6854}\inst{\ref{aff70}}
\and W.~Roster\orcid{0000-0002-9149-6528}\inst{\ref{aff50}}
\and S.~Sacquegna\orcid{0000-0002-8433-6630}\inst{\ref{aff148}}
\and M.~Sahl\'en\orcid{0000-0003-0973-4804}\inst{\ref{aff149}}
\and D.~B.~Sanders\orcid{0000-0002-1233-9998}\inst{\ref{aff33}}
\and E.~Sarpa\orcid{0000-0002-1256-655X}\inst{\ref{aff14},\ref{aff113},\ref{aff12}}
\and C.~Scarlata\orcid{0000-0002-9136-8876}\inst{\ref{aff150}}
\and A.~Schneider\orcid{0000-0001-7055-8104}\inst{\ref{aff142}}
\and D.~Sciotti\orcid{0009-0008-4519-2620}\inst{\ref{aff31},\ref{aff72}}
\and E.~Sellentin\inst{\ref{aff151},\ref{aff25}}
\and L.~C.~Smith\orcid{0000-0002-3259-2771}\inst{\ref{aff152}}
\and J.~G.~Sorce\orcid{0000-0002-2307-2432}\inst{\ref{aff153},\ref{aff45}}
\and F.~Tarsitano\orcid{0000-0002-5919-0238}\inst{\ref{aff154},\ref{aff44}}
\and G.~Testera\inst{\ref{aff18}}
\and R.~Teyssier\orcid{0000-0001-7689-0933}\inst{\ref{aff155}}
\and S.~Tosi\orcid{0000-0002-7275-9193}\inst{\ref{aff9},\ref{aff18},\ref{aff17}}
\and A.~Troja\orcid{0000-0003-0239-4595}\inst{\ref{aff90},\ref{aff91}}
\and A.~Venhola\orcid{0000-0001-6071-4564}\inst{\ref{aff156}}
\and D.~Vergani\orcid{0000-0003-0898-2216}\inst{\ref{aff10}}
\and G.~Verza\orcid{0000-0002-1886-8348}\inst{\ref{aff157},\ref{aff158}}
\and S.~Vinciguerra\orcid{0009-0005-4018-3184}\inst{\ref{aff73}}
\and N.~A.~Walton\orcid{0000-0003-3983-8778}\inst{\ref{aff152}}
\and J.~R.~Weaver\orcid{0000-0003-1614-196X}\inst{\ref{aff159}}
\and A.~H.~Wright\orcid{0000-0001-7363-7932}\inst{\ref{aff135}}}

\institute{Department of Physics \& Astronomy, University of California Irvine, Irvine CA 92697, USA\label{aff1}
\and
California Institute of Technology, 1200 E California Blvd, Pasadena, CA 91125, USA\label{aff2}
\and
Center for Astrophysics and Cosmology, University of Nova Gorica, Nova Gorica, Slovenia\label{aff3}
\and
Kavli Institute for Cosmology Cambridge, Madingley Road, Cambridge, CB3 0HA, UK\label{aff4}
\and
Cavendish Laboratory, University of Cambridge, JJ Thomson Avenue, Cambridge, CB3 0HE, UK\label{aff5}
\and
Department of Physics and Astronomy, University of British Columbia, Vancouver, BC V6T 1Z1, Canada\label{aff6}
\and
ESAC/ESA, Camino Bajo del Castillo, s/n., Urb. Villafranca del Castillo, 28692 Villanueva de la Ca\~nada, Madrid, Spain\label{aff7}
\and
School of Mathematics and Physics, University of Surrey, Guildford, Surrey, GU2 7XH, UK\label{aff8}
\and
INAF-Osservatorio Astronomico di Brera, Via Brera 28, 20122 Milano, Italy\label{aff9}
\and
INAF-Osservatorio di Astrofisica e Scienza dello Spazio di Bologna, Via Piero Gobetti 93/3, 40129 Bologna, Italy\label{aff10}
\and
IFPU, Institute for Fundamental Physics of the Universe, via Beirut 2, 34151 Trieste, Italy\label{aff11}
\and
INAF-Osservatorio Astronomico di Trieste, Via G. B. Tiepolo 11, 34143 Trieste, Italy\label{aff12}
\and
INFN, Sezione di Trieste, Via Valerio 2, 34127 Trieste TS, Italy\label{aff13}
\and
SISSA, International School for Advanced Studies, Via Bonomea 265, 34136 Trieste TS, Italy\label{aff14}
\and
Dipartimento di Fisica e Astronomia, Universit\`a di Bologna, Via Gobetti 93/2, 40129 Bologna, Italy\label{aff15}
\and
INFN-Sezione di Bologna, Viale Berti Pichat 6/2, 40127 Bologna, Italy\label{aff16}
\and
Dipartimento di Fisica, Universit\`a di Genova, Via Dodecaneso 33, 16146, Genova, Italy\label{aff17}
\and
INFN-Sezione di Genova, Via Dodecaneso 33, 16146, Genova, Italy\label{aff18}
\and
Department of Physics "E. Pancini", University Federico II, Via Cinthia 6, 80126, Napoli, Italy\label{aff19}
\and
INAF-Osservatorio Astronomico di Capodimonte, Via Moiariello 16, 80131 Napoli, Italy\label{aff20}
\and
Dipartimento di Fisica, Universit\`a degli Studi di Torino, Via P. Giuria 1, 10125 Torino, Italy\label{aff21}
\and
INFN-Sezione di Torino, Via P. Giuria 1, 10125 Torino, Italy\label{aff22}
\and
INAF-Osservatorio Astrofisico di Torino, Via Osservatorio 20, 10025 Pino Torinese (TO), Italy\label{aff23}
\and
Institute for Astronomy, University of Edinburgh, Royal Observatory, Blackford Hill, Edinburgh EH9 3HJ, UK\label{aff24}
\and
Leiden Observatory, Leiden University, Einsteinweg 55, 2333 CC Leiden, The Netherlands\label{aff25}
\and
INAF-IASF Milano, Via Alfonso Corti 12, 20133 Milano, Italy\label{aff26}
\and
Centro de Investigaciones Energ\'eticas, Medioambientales y Tecnol\'ogicas (CIEMAT), Avenida Complutense 40, 28040 Madrid, Spain\label{aff27}
\and
Port d'Informaci\'{o} Cient\'{i}fica, Campus UAB, C. Albareda s/n, 08193 Bellaterra (Barcelona), Spain\label{aff28}
\and
Institute for Theoretical Particle Physics and Cosmology (TTK), RWTH Aachen University, 52056 Aachen, Germany\label{aff29}
\and
Deutsches Zentrum f\"ur Luft- und Raumfahrt e. V. (DLR), Linder H\"ohe, 51147 K\"oln, Germany\label{aff30}
\and
INAF-Osservatorio Astronomico di Roma, Via Frascati 33, 00078 Monteporzio Catone, Italy\label{aff31}
\and
INFN section of Naples, Via Cinthia 6, 80126, Napoli, Italy\label{aff32}
\and
Institute for Astronomy, University of Hawaii, 2680 Woodlawn Drive, Honolulu, HI 96822, USA\label{aff33}
\and
Dipartimento di Fisica e Astronomia "Augusto Righi" - Alma Mater Studiorum Universit\`a di Bologna, Viale Berti Pichat 6/2, 40127 Bologna, Italy\label{aff34}
\and
Instituto de Astrof\'{\i}sica de Canarias, E-38205 La Laguna, Tenerife, Spain\label{aff35}
\and
Jodrell Bank Centre for Astrophysics, Department of Physics and Astronomy, University of Manchester, Oxford Road, Manchester M13 9PL, UK\label{aff36}
\and
European Space Agency/ESRIN, Largo Galileo Galilei 1, 00044 Frascati, Roma, Italy\label{aff37}
\and
Universit\'e Claude Bernard Lyon 1, CNRS/IN2P3, IP2I Lyon, UMR 5822, Villeurbanne, F-69100, France\label{aff38}
\and
Institut de Ci\`{e}ncies del Cosmos (ICCUB), Universitat de Barcelona (IEEC-UB), Mart\'{i} i Franqu\`{e}s 1, 08028 Barcelona, Spain\label{aff39}
\and
Instituci\'o Catalana de Recerca i Estudis Avan\c{c}ats (ICREA), Passeig de Llu\'{\i}s Companys 23, 08010 Barcelona, Spain\label{aff40}
\and
Institut de Ciencies de l'Espai (IEEC-CSIC), Campus UAB, Carrer de Can Magrans, s/n Cerdanyola del Vall\'es, 08193 Barcelona, Spain\label{aff41}
\and
UCB Lyon 1, CNRS/IN2P3, IUF, IP2I Lyon, 4 rue Enrico Fermi, 69622 Villeurbanne, France\label{aff42}
\and
Universit\'e Paris-Saclay, Universit\'e Paris Cit\'e, CEA, CNRS, AIM, 91191, Gif-sur-Yvette, France\label{aff43}
\and
Department of Astronomy, University of Geneva, ch. d'Ecogia 16, 1290 Versoix, Switzerland\label{aff44}
\and
Universit\'e Paris-Saclay, CNRS, Institut d'astrophysique spatiale, 91405, Orsay, France\label{aff45}
\and
INAF-Istituto di Astrofisica e Planetologia Spaziali, via del Fosso del Cavaliere, 100, 00100 Roma, Italy\label{aff46}
\and
Space Science Data Center, Italian Space Agency, via del Politecnico snc, 00133 Roma, Italy\label{aff47}
\and
INFN-Bologna, Via Irnerio 46, 40126 Bologna, Italy\label{aff48}
\and
University Observatory, LMU Faculty of Physics, Scheinerstr.~1, 81679 Munich, Germany\label{aff49}
\and
Max Planck Institute for Extraterrestrial Physics, Giessenbachstr. 1, 85748 Garching, Germany\label{aff50}
\and
INAF-Osservatorio Astronomico di Padova, Via dell'Osservatorio 5, 35122 Padova, Italy\label{aff51}
\and
Universit\"ats-Sternwarte M\"unchen, Fakult\"at f\"ur Physik, Ludwig-Maximilians-Universit\"at M\"unchen, Scheinerstr.~1, 81679 M\"unchen, Germany\label{aff52}
\and
Institute of Theoretical Astrophysics, University of Oslo, P.O. Box 1029 Blindern, 0315 Oslo, Norway\label{aff53}
\and
Jet Propulsion Laboratory, California Institute of Technology, 4800 Oak Grove Drive, Pasadena, CA, 91109, USA\label{aff54}
\and
Felix Hormuth Engineering, Goethestr. 17, 69181 Leimen, Germany\label{aff55}
\and
Technical University of Denmark, Elektrovej 327, 2800 Kgs. Lyngby, Denmark\label{aff56}
\and
Cosmic Dawn Center (DAWN), Denmark\label{aff57}
\and
Max-Planck-Institut f\"ur Astronomie, K\"onigstuhl 17, 69117 Heidelberg, Germany\label{aff58}
\and
NASA Goddard Space Flight Center, Greenbelt, MD 20771, USA\label{aff59}
\and
Department of Physics and Astronomy, University College London, Gower Street, London WC1E 6BT, UK\label{aff60}
\and
Aix-Marseille Universit\'e, CNRS/IN2P3, CPPM, Marseille, France\label{aff61}
\and
Universit\'e de Gen\`eve, D\'epartement de Physique Th\'eorique and Centre for Astroparticle Physics, 24 quai Ernest-Ansermet, CH-1211 Gen\`eve 4, Switzerland\label{aff62}
\and
Department of Physics, P.O. Box 64, University of Helsinki, 00014 Helsinki, Finland\label{aff63}
\and
Helsinki Institute of Physics, Gustaf H{\"a}llstr{\"o}min katu 2, University of Helsinki, 00014 Helsinki, Finland\label{aff64}
\and
Laboratoire d'etude de l'Univers et des phenomenes eXtremes, Observatoire de Paris, Universit\'e PSL, Sorbonne Universit\'e, CNRS, 92190 Meudon, France\label{aff65}
\and
SKAO, Jodrell Bank, Lower Withington, Macclesfield SK11 9FT, UK\label{aff66}
\and
Centre de Calcul de l'IN2P3/CNRS, 21 avenue Pierre de Coubertin 69627 Villeurbanne Cedex, France\label{aff67}
\and
Dipartimento di Fisica "Aldo Pontremoli", Universit\`a degli Studi di Milano, Via Celoria 16, 20133 Milano, Italy\label{aff68}
\and
INFN-Sezione di Milano, Via Celoria 16, 20133 Milano, Italy\label{aff69}
\and
University of Applied Sciences and Arts of Northwestern Switzerland, School of Computer Science, 5210 Windisch, Switzerland\label{aff70}
\and
Universit\"at Bonn, Argelander-Institut f\"ur Astronomie, Auf dem H\"ugel 71, 53121 Bonn, Germany\label{aff71}
\and
INFN-Sezione di Roma, Piazzale Aldo Moro, 2 - c/o Dipartimento di Fisica, Edificio G. Marconi, 00185 Roma, Italy\label{aff72}
\and
Aix-Marseille Universit\'e, CNRS, CNES, LAM, Marseille, France\label{aff73}
\and
Dipartimento di Fisica e Astronomia "Augusto Righi" - Alma Mater Studiorum Universit\`a di Bologna, via Piero Gobetti 93/2, 40129 Bologna, Italy\label{aff74}
\and
Department of Physics, Institute for Computational Cosmology, Durham University, South Road, Durham, DH1 3LE, UK\label{aff75}
\and
Universit\'e Paris Cit\'e, CNRS, Astroparticule et Cosmologie, 75013 Paris, France\label{aff76}
\and
CNRS-UCB International Research Laboratory, Centre Pierre Bin\'etruy, IRL2007, CPB-IN2P3, Berkeley, USA\label{aff77}
\and
Institut d'Astrophysique de Paris, 98bis Boulevard Arago, 75014, Paris, France\label{aff78}
\and
Institut d'Astrophysique de Paris, UMR 7095, CNRS, and Sorbonne Universit\'e, 98 bis boulevard Arago, 75014 Paris, France\label{aff79}
\and
Institute of Physics, Laboratory of Astrophysics, Ecole Polytechnique F\'ed\'erale de Lausanne (EPFL), Observatoire de Sauverny, 1290 Versoix, Switzerland\label{aff80}
\and
Telespazio UK S.L. for European Space Agency (ESA), Camino bajo del Castillo, s/n, Urbanizacion Villafranca del Castillo, Villanueva de la Ca\~nada, 28692 Madrid, Spain\label{aff81}
\and
Institut de F\'{i}sica d'Altes Energies (IFAE), The Barcelona Institute of Science and Technology, Campus UAB, 08193 Bellaterra (Barcelona), Spain\label{aff82}
\and
European Space Agency/ESTEC, Keplerlaan 1, 2201 AZ Noordwijk, The Netherlands\label{aff83}
\and
DARK, Niels Bohr Institute, University of Copenhagen, Jagtvej 155, 2200 Copenhagen, Denmark\label{aff84}
\and
Waterloo Centre for Astrophysics, University of Waterloo, Waterloo, Ontario N2L 3G1, Canada\label{aff85}
\and
Department of Physics and Astronomy, University of Waterloo, Waterloo, Ontario N2L 3G1, Canada\label{aff86}
\and
Perimeter Institute for Theoretical Physics, Waterloo, Ontario N2L 2Y5, Canada\label{aff87}
\and
Centre National d'Etudes Spatiales -- Centre spatial de Toulouse, 18 avenue Edouard Belin, 31401 Toulouse Cedex 9, France\label{aff88}
\and
Institute of Space Science, Str. Atomistilor, nr. 409 M\u{a}gurele, Ilfov, 077125, Romania\label{aff89}
\and
Dipartimento di Fisica e Astronomia "G. Galilei", Universit\`a di Padova, Via Marzolo 8, 35131 Padova, Italy\label{aff90}
\and
INFN-Padova, Via Marzolo 8, 35131 Padova, Italy\label{aff91}
\and
Institut f\"ur Theoretische Physik, University of Heidelberg, Philosophenweg 16, 69120 Heidelberg, Germany\label{aff92}
\and
Institut de Recherche en Astrophysique et Plan\'etologie (IRAP), Universit\'e de Toulouse, CNRS, UPS, CNES, 14 Av. Edouard Belin, 31400 Toulouse, France\label{aff93}
\and
Universit\'e St Joseph; Faculty of Sciences, Beirut, Lebanon\label{aff94}
\and
Departamento de F\'isica, FCFM, Universidad de Chile, Blanco Encalada 2008, Santiago, Chile\label{aff95}
\and
Universit\"at Innsbruck, Institut f\"ur Astro- und Teilchenphysik, Technikerstr. 25/8, 6020 Innsbruck, Austria\label{aff96}
\and
Institut d'Estudis Espacials de Catalunya (IEEC),  Edifici RDIT, Campus UPC, 08860 Castelldefels, Barcelona, Spain\label{aff97}
\and
Satlantis, University Science Park, Sede Bld 48940, Leioa-Bilbao, Spain\label{aff98}
\and
Institute of Space Sciences (ICE, CSIC), Campus UAB, Carrer de Can Magrans, s/n, 08193 Barcelona, Spain\label{aff99}
\and
Department of Physics and Helsinki Institute of Physics, Gustaf H\"allstr\"omin katu 2, University of Helsinki, 00014 Helsinki, Finland\label{aff100}
\and
Departamento de F\'isica, Faculdade de Ci\^encias, Universidade de Lisboa, Edif\'icio C8, Campo Grande, PT1749-016 Lisboa, Portugal\label{aff101}
\and
Instituto de Astrof\'isica e Ci\^encias do Espa\c{c}o, Faculdade de Ci\^encias, Universidade de Lisboa, Tapada da Ajuda, 1349-018 Lisboa, Portugal\label{aff102}
\and
Mullard Space Science Laboratory, University College London, Holmbury St Mary, Dorking, Surrey RH5 6NT, UK\label{aff103}
\and
Cosmic Dawn Center (DAWN)\label{aff104}
\and
Niels Bohr Institute, University of Copenhagen, Jagtvej 128, 2200 Copenhagen, Denmark\label{aff105}
\and
Universidad Polit\'ecnica de Cartagena, Departamento de Electr\'onica y Tecnolog\'ia de Computadoras,  Plaza del Hospital 1, 30202 Cartagena, Spain\label{aff106}
\and
Caltech/IPAC, 1200 E. California Blvd., Pasadena, CA 91125, USA\label{aff107}
\and
Dipartimento di Fisica e Scienze della Terra, Universit\`a degli Studi di Ferrara, Via Giuseppe Saragat 1, 44122 Ferrara, Italy\label{aff108}
\and
Istituto Nazionale di Fisica Nucleare, Sezione di Ferrara, Via Giuseppe Saragat 1, 44122 Ferrara, Italy\label{aff109}
\and
INAF, Istituto di Radioastronomia, Via Piero Gobetti 101, 40129 Bologna, Italy\label{aff110}
\and
Astronomical Observatory of the Autonomous Region of the Aosta Valley (OAVdA), Loc. Lignan 39, I-11020, Nus (Aosta Valley), Italy\label{aff111}
\and
Universit\'e C\^{o}te d'Azur, Observatoire de la C\^{o}te d'Azur, CNRS, Laboratoire Lagrange, Bd de l'Observatoire, CS 34229, 06304 Nice cedex 4, France\label{aff112}
\and
ICSC - Centro Nazionale di Ricerca in High Performance Computing, Big Data e Quantum Computing, Via Magnanelli 2, Bologna, Italy\label{aff113}
\and
Department of Physics, Oxford University, Keble Road, Oxford OX1 3RH, UK\label{aff114}
\and
Univ. Grenoble Alpes, CNRS, Grenoble INP, LPSC-IN2P3, 53, Avenue des Martyrs, 38000, Grenoble, France\label{aff115}
\and
Dipartimento di Fisica, Sapienza Universit\`a di Roma, Piazzale Aldo Moro 2, 00185 Roma, Italy\label{aff116}
\and
Aurora Technology for European Space Agency (ESA), Camino bajo del Castillo, s/n, Urbanizacion Villafranca del Castillo, Villanueva de la Ca\~nada, 28692 Madrid, Spain\label{aff117}
\and
Dipartimento di Fisica - Sezione di Astronomia, Universit\`a di Trieste, Via Tiepolo 11, 34131 Trieste, Italy\label{aff118}
\and
Department of Mathematics and Physics E. De Giorgi, University of Salento, Via per Arnesano, CP-I93, 73100, Lecce, Italy\label{aff119}
\and
INFN, Sezione di Lecce, Via per Arnesano, CP-193, 73100, Lecce, Italy\label{aff120}
\and
INAF-Sezione di Lecce, c/o Dipartimento Matematica e Fisica, Via per Arnesano, 73100, Lecce, Italy\label{aff121}
\and
ICL, Junia, Universit\'e Catholique de Lille, LITL, 59000 Lille, France\label{aff122}
\and
Instituto de F\'isica Te\'orica UAM-CSIC, Campus de Cantoblanco, 28049 Madrid, Spain\label{aff123}
\and
CERCA/ISO, Department of Physics, Case Western Reserve University, 10900 Euclid Avenue, Cleveland, OH 44106, USA\label{aff124}
\and
Laboratoire Univers et Th\'eorie, Observatoire de Paris, Universit\'e PSL, Universit\'e Paris Cit\'e, CNRS, 92190 Meudon, France\label{aff125}
\and
IRFU, CEA, Universit\'e Paris-Saclay 91191 Gif-sur-Yvette Cedex, France\label{aff126}
\and
Universit\'e de Strasbourg, CNRS, Observatoire astronomique de Strasbourg, UMR 7550, 67000 Strasbourg, France\label{aff127}
\and
Center for Data-Driven Discovery, Kavli IPMU (WPI), UTIAS, The University of Tokyo, Kashiwa, Chiba 277-8583, Japan\label{aff128}
\and
Kapteyn Astronomical Institute, University of Groningen, PO Box 800, 9700 AV Groningen, The Netherlands\label{aff129}
\and
Departamento F\'isica Aplicada, Universidad Polit\'ecnica de Cartagena, Campus Muralla del Mar, 30202 Cartagena, Murcia, Spain\label{aff130}
\and
Instituto de F\'isica de Cantabria, Edificio Juan Jord\'a, Avenida de los Castros, 39005 Santander, Spain\label{aff131}
\and
Institute of Cosmology and Gravitation, University of Portsmouth, Portsmouth PO1 3FX, UK\label{aff132}
\and
Department of Computer Science, Aalto University, PO Box 15400, Espoo, FI-00 076, Finland\label{aff133}
\and
Universidad de La Laguna, Dpto. Astrof\'\i sica, E-38206 La Laguna, Tenerife, Spain\label{aff134}
\and
Ruhr University Bochum, Faculty of Physics and Astronomy, Astronomical Institute (AIRUB), German Centre for Cosmological Lensing (GCCL), 44780 Bochum, Germany\label{aff135}
\and
Department of Physics and Astronomy, Vesilinnantie 5, University of Turku, 20014 Turku, Finland\label{aff136}
\and
Serco for European Space Agency (ESA), Camino bajo del Castillo, s/n, Urbanizacion Villafranca del Castillo, Villanueva de la Ca\~nada, 28692 Madrid, Spain\label{aff137}
\and
ARC Centre of Excellence for Dark Matter Particle Physics, Melbourne, Australia\label{aff138}
\and
Centre for Astrophysics \& Supercomputing, Swinburne University of Technology,  Hawthorn, Victoria 3122, Australia\label{aff139}
\and
Department of Physics and Astronomy, University of the Western Cape, Bellville, Cape Town, 7535, South Africa\label{aff140}
\and
DAMTP, Centre for Mathematical Sciences, Wilberforce Road, Cambridge CB3 0WA, UK\label{aff141}
\and
Department of Astrophysics, University of Zurich, Winterthurerstrasse 190, 8057 Zurich, Switzerland\label{aff142}
\and
Department of Physics, Centre for Extragalactic Astronomy, Durham University, South Road, Durham, DH1 3LE, UK\label{aff143}
\and
INAF-Osservatorio Astrofisico di Arcetri, Largo E. Fermi 5, 50125, Firenze, Italy\label{aff144}
\and
Centro de Astrof\'{\i}sica da Universidade do Porto, Rua das Estrelas, 4150-762 Porto, Portugal\label{aff145}
\and
Instituto de Astrof\'isica e Ci\^encias do Espa\c{c}o, Universidade do Porto, CAUP, Rua das Estrelas, PT4150-762 Porto, Portugal\label{aff146}
\and
HE Space for European Space Agency (ESA), Camino bajo del Castillo, s/n, Urbanizacion Villafranca del Castillo, Villanueva de la Ca\~nada, 28692 Madrid, Spain\label{aff147}
\and
INAF - Osservatorio Astronomico d'Abruzzo, Via Maggini, 64100, Teramo, Italy\label{aff148}
\and
Theoretical astrophysics, Department of Physics and Astronomy, Uppsala University, Box 516, 751 37 Uppsala, Sweden\label{aff149}
\and
Minnesota Institute for Astrophysics, University of Minnesota, 116 Church St SE, Minneapolis, MN 55455, USA\label{aff150}
\and
Mathematical Institute, University of Leiden, Einsteinweg 55, 2333 CA Leiden, The Netherlands\label{aff151}
\and
Institute of Astronomy, University of Cambridge, Madingley Road, Cambridge CB3 0HA, UK\label{aff152}
\and
Univ. Lille, CNRS, Centrale Lille, UMR 9189 CRIStAL, 59000 Lille, France\label{aff153}
\and
Institute for Particle Physics and Astrophysics, Dept. of Physics, ETH Zurich, Wolfgang-Pauli-Strasse 27, 8093 Zurich, Switzerland\label{aff154}
\and
Department of Astrophysical Sciences, Peyton Hall, Princeton University, Princeton, NJ 08544, USA\label{aff155}
\and
Space physics and astronomy research unit, University of Oulu, Pentti Kaiteran katu 1, FI-90014 Oulu, Finland\label{aff156}
\and
International Centre for Theoretical Physics (ICTP), Strada Costiera 11, 34151 Trieste, Italy\label{aff157}
\and
Center for Computational Astrophysics, Flatiron Institute, 162 5th Avenue, 10010, New York, NY, USA\label{aff158}
\and
MIT Kavli Institute for Astrophysics and Space Research, Massachusetts Institute of Technology, Cambridge, MA 02139, USA\label{aff159}}    

\abstract{The extragalactic background light (EBL) fluctuations in the optical/near-IR encode the cumulative emission of unresolved galaxies, integrated galaxy light (IGL), diffuse intra-halo light (IHL), and high-$z$ sources from the epoch of reionisation (EoR), but they are difficult to disentangle with auto-spectra alone. Our aim was to decompose the EBL into its principal constituents using multi-band intensity mapping combined with cosmic shear and galaxy clustering. We developed a joint halo-model framework in which IHL follows a mass- and redshift-dependent luminosity scaling, IGL is set by an evolving Schechter luminosity function, and EoR emission is modelled with Pop II/III stellar emissivities and a binned star formation efficiency. Cosmic shear is modelled using the tidal alignment and tidal torquing model for intrinsic alignment. Using mock surveys in a flat Lambda cold dark matter ($\Lambda$CDM) cosmology with ten spectral bands spanning 0.75--5.0\,$\si{\micron}$ in the north ecliptic pole deep fields over about $100\,\deg^2$ with source detections down to AB\,=\,20.5 for masking, and six redshift bins to $z=2.5$, we fit auto- and cross-power spectra using a Markov chain Monte Carlo method. The combined SPHEREx$\times$\Euclid analysis recovers all fiducial parameters within $1\,\sigma$ and reduces $1\,\sigma$ uncertainties on IHL parameters by 10--30\% relative to SPHEREx EBL-only, while EoR star formation efficiency parameters improve by 20--30\%. The predicted cross-correlations show a stronger coupling of IHL than IGL to the shear field within adopted model framework, enhancing component separation; conversely, the high-$z$ EoR contribution shows negligible correlation with cosmic shear and galaxy clustering, aiding its isolation in the EBL. Relative to the SPHEREx EBL-only case, the inferred IHL fraction as a function of halo mass is significantly tightened over $10^{11}$–$10^{14}\,\si{\solarmass}$, with uncertainties reduced by 5--30\%, and the resulting star formation rate density constraints extend to $z\approx11$, with uncertainty reductions of 15--30\%. SPHEREx$\times$\Euclid provides a robust systematics-aware route to component-resolved EBL measurements and improved constraints on galaxy formation.}
\titlerunning{\Euclid\/: SPHEREx $\times$ \Euclid}
\authorrunning{Euclid Collaboration: Cao et al.}
\maketitle
\nolinenumbers

\section{Introduction}
The optical and near-infrared extragalactic background light (EBL) encodes the cumulative radiation from all sources over cosmic time. Its absolute intensity level and spatial fluctuations provide a uniquely integrated view of galaxy formation and the build-up of stellar mass and dust across environments and redshifts \citep{Kashlinsky05,Dole06}. Multiple components contribute to the EBL anisotropy: diffuse intra-halo light (IHL) from tidally stripped and intra-group and/or cluster stars (e.g. \citealt{Cooray12b,Zemcov14}); unresolved integrated galaxy light (IGL) from faint galaxies below the detection threshold; and high-redshift emission from the epoch of reionisation (EoR). In addition, there are foregrounds such as diffuse Galactic light (DGL) from dust scattering and interplanetary dust (zodiacal light), as well as instrument noise and Poisson noise from discrete sources. The central challenge is that these signals overlap in angular scales and, in broad bands, in spectral colour, making the auto-power spectra of intensity maps alone insufficient to cleanly decompose them.

Historically, the EBL has been constrained by both absolute measurements and fluctuation analyses. Early attempts \citep{Dicke65} were followed by the first robust far-infrared detections with \textit{COBE}/DIRBE and \textit{FIRAS} \citep{Puget96,Fixsen98}. More recently, the combination of higher sensitivity and improved angular resolution has enabled precise fluctuation measurements and deeper masking, from facilities spanning \textit{Spitzer} and ground-based telescopes. Looking forward, the Spectro-Photometer for the History of the Universe, Epoch of Reionization, and Ices Explorer (SPHEREx; \citealt{Dore14,Bock26}) will deliver an all-sky spectral survey with unprecedented multi-band coverage in the near-IR. Its deep fields overlap with wide-field optical imaging and lensing programmes, providing exactly the multi-probe context needed to isolate EBL components. 

In a previous study the internal multi-band cross-power spectra of SPHEREx data alone were considered as a way to decompose principal components of the EBL \citep{Feng19}. Here, we consider the added information provided by external tracer fields. In particular, weak gravitational lensing (cosmic shear) has matured into a leading cosmological probe of dark matter clustering and cosmic acceleration \citep{Kaiser92,Refregier03,Albrecht06,Peacock06,Mandelbaum18,Blake20}. The \Euclid mission \citep{EuclidSkyOverview} will map shapes and photometric redshifts for billions of galaxies over about 14\,000\,$\deg^2$, enabling high-precision tomographic shear and galaxy-clustering measurements \citep{Laureijs11}. Additional surveys such as the China Space Station Telescope (CSST; \citealt{Zhan11,Cao18}) and the Vera C.\ Rubin Observatory Legacy Survey of Space and Time (LSST; \citealt{LSST09,Ivezic19}) will further enrich the available data.

This paper exploits the complementarity between optical/IR intensity maps with cosmic shear and galaxy clustering. The key idea is that different EBL contributors trace the large-scale structure with distinct (bias-weighted) redshift kernels, and therefore their cross-correlations with shear and with galaxy density carry diagnostic power that auto-spectra lack. IHL, likely originating from tidally stripped stars, merger debris, and other diffuse stellar components within haloes, is treated as a diffuse stellar component associated with dark matter haloes \citep{Purcell07,Cooray12}. By construction, it follows the distribution of stars bound to haloes and hence correlates strongly with the gravitational field traced by cosmic shear maps. IGL, built from galaxies fainter than the masking threshold, also correlates with structure, but with a different mass and redshift weighting set by the evolving luminosity function and halo occupation \citep{Lagache03,Dole04,Lim23}. By contrast, the high-redshift EoR component has little overlap with the low-redshift kernels of \Euclid shear and clustering, and its cross-signal is therefore strongly suppressed. However, this contrast helps isolate the EoR contribution once the low-redshift components are better constrained by the cross-correlations. These differences, when combined with SPHEREx’s ten spectrally cladded  broad bands spanning 0.75--5.0\,$\si{\micron}$ in the deep fields, provide the basis for a multi-band, multi-probe approach to separate the IHL, IGL, and EoR contributions.

Cross-correlation between different datasets also mitigates foregrounds and systematic effects. Zodiacal fluctuations and instrumental noise in the SPHEREx intensity maps are largely uncorrelated with the overlapping Euclid weak lensing and galaxy clustering data, and thus do not bias the mean cross-power. DGL, which follows Milky Way dust, is a major foreground in the EBL auto-spectra on large angular scales, typically at angular scales of tens of arcmin and above \citep{Mitchell15}, but is likely to have a minimal correlation with extragalactic shear fields at high Galactic latitude. On the lensing side, multiplicative shear calibration and intrinsic alignments (IA; \citealt{Hirata04,Troxel15,Joachim15,Blazek19}) are the leading systematic effects; they can be modelled and marginalised within a tidal alignment and tidal torquing (TATT; \citealt{Blazek19}) framework while retaining most of the cosmological and astrophysical information content. In short, cross-spectra provide both a lever arm for component separation and a robust pathway through the dominant single-probe systematic effects.

For this paper we developed a forward model that captures these ingredients with minimal complexity while remaining physically interpretable. For IHL we adopted a halo-model description in which the IHL luminosity scales with halo mass and evolves with redshift via a power law; we then allowed the mass dependence and redshift evolution to vary in tomographic bins to test for departures from simple self-similarity \citep{Cooray12}. For IGL, we integrated a redshift-evolving Schechter luminosity function \citep{Helgason12} to compute number counts and flux production below the masking threshold and to connect those galaxies to haloes through a halo occupation distribution (HOD) prescription. For EoR, we used template emissivities for Pop~II/III stellar populations and parametrised the star formation efficiency in redshift bins \citep{Fernandez06,Fernandez10,Fernandez12}, which directly maps to the contribution of reionising sources in the SPHEREx bands. DGL was modelled as a power law in multipole at high latitude, and we included a shot-noise term whose level depends on the effective magnitude cut. On the shear side we modelled tomographic $C_\ell$s with Limber projection \citep{Limber53,LoVerde08}, included IA through the TATT model, and adopted per-bin Gaussian priors on multiplicative shear calibration motivated by stage IV survey requirements. Galaxy clustering enters through tomographic auto-spectra. The contribution of cross-bin clustering power spectra is not dominant in this analysis and does not significantly improve the EBL constraints; therefore, we neglected the correlations between different tomographic bins for galaxy clustering.

The SPHEREx$\times$\Euclid synergy is especially powerful because the two experiments bring complementary strengths in angular and spectral coverage. SPHEREx provides many narrow bands with modest angular resolution but deep, uniform coverage in the north ecliptic pole (NEP) and the south ecliptic pole (SEP), enabling precise multi-band EBL auto- and cross-spectra. At high $\ell$, once instrumental noise and shot noise are controlled, these measurements are shape-noise and shot-noise dominated. \Euclid contributes high signal-to-noise shear maps in several photometric redshift bins and dense galaxy samples for clustering over two deep fields, the Euclid Deep Field North (EDFN; 20\,$\deg^2$) and the Euclid Deep Field South (EDFS; 23\,$\deg^2$), both of which are fully contained within the corresponding SPHEREx deep fields \citep{Dore18,Scaramella-EP1}. The joint data vector thus contains ten EBL auto-spectra, tomographic shear and galaxy auto-spectra, and crucially cross-spectra between each EBL band and each shear (and clustering) bin. The spectral diversity across the ten bands constrains the colour of each component; the tomographic information in shear and galaxies constrains its redshift weighting; and the cross-correlation pattern across probes breaks degeneracies between components that project similarly in any one map alone.

In this paper we present end-to-end forecasts based on mock surveys closely matched to SPHEREx deep-field imaging (ten bands between 0.75 and 5.0\,$\si{\micron}$; AB\,=\,20.5 masking at 2.2\,$\si{\micron}$) and \Euclid tomography (six shear--clustering bins to $z=2.5$). We compute auto- and cross-power spectra for EBL, shear, and galaxies, including realistic noise terms, and fit a multi-component halo-model with a Markov chain Monte Carlo (MCMC) approach. Our goals are twofold: (1) to quantify how cross-correlations tighten constraints on the IHL mass and redshift scaling and on the EoR star formation efficiency relative to EBL-only analyses; and (2) to assess the extent to which cross-spectra suppress or diagnose key systematic effects (DGL, shot noise, shear calibration, and IA). Our combined analysis recovers the input parameters within $1\,\sigma$ and reduces marginalised uncertainties on IHL parameters by 10--30\% and on EoR star formation efficiency by 20--30\% compared to SPHEREx-only fits, while tightening the inferred IHL fraction across $10^{11.2}$--$10^{13.8}\,\si{\solarmass} $ and extending star formation rate density constraints to $z\approx11$, by better isolating the low-redshift EBL components.

The remainder of the paper is organised as follows. In Sect.\,\ref{sc:model} we present the theoretical framework for EBL components, cosmic shear, and galaxy clustering.  In Sect.\,\ref{analysis} we introduce the instrument parameters of SPHEREx and \Euclid, and the method used to create the mock data. We then present the measured power spectra and use a multi-component model to fit them using the MCMC method. In Sect.\,\ref{discussion} we discuss the stability of our results and the limitations of our methodology. We summarise our main findings and discuss implications for component-resolved EBL science and early-galaxy constraints in Sect.\,\ref{conclusions}. In this work we adopt the standard Lambda cold dark matter ($\Lambda$CDM) cosmological model with parameter values $\si{\hubble} =0.67$, $\Omega_{\rm m} = 0.31$, $\Omega_{\rm b} = 0.05$, $\sigma_8=0.81$, and $n_{\rm s} = 0.96$ \citep{Aghanim20}.

\section{\label{sc:model}Theoretical model}
\begin{figure*}
\centering
\includegraphics[width=0.95\textwidth]{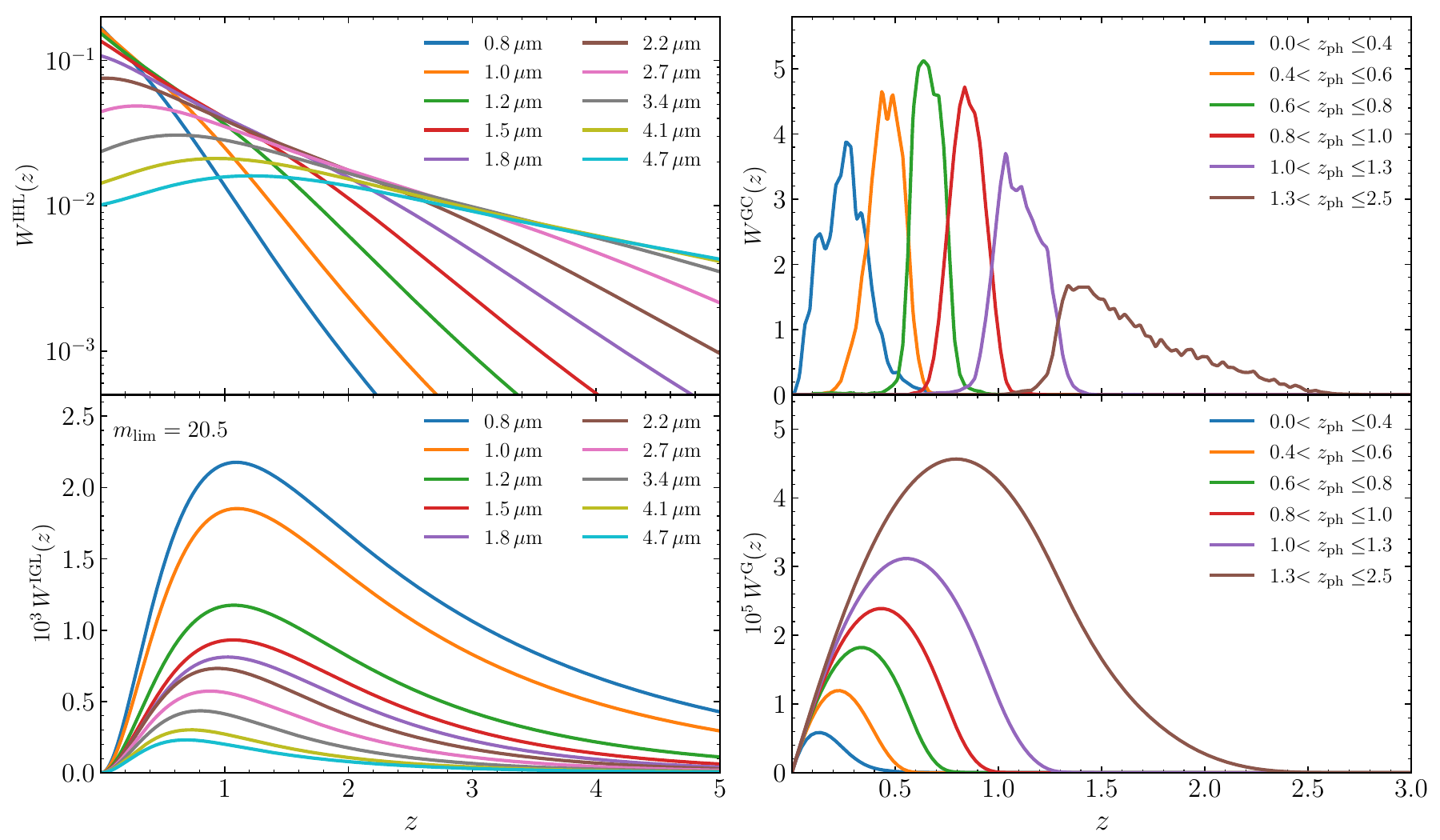}
\caption{Kernels of different EBL components and cosmic shear signal as functions of redshift. \emph{Left}: Kernels of IHL (\emph{upper}) and IGL (\emph{lower}) signals in EBL (IGL with masking down to AB\,=\,20.5). \emph{Right}: Kernels of the galaxy clustering (\emph{upper}) and cosmic shear (\emph{lower}).}
\label{fig:kernel}
\end{figure*}

For a flat universe, we can calculate the angular power spectrum using the Limber approximation \citep{Limber53,LoVerde08},
\begin{equation}
C_{XY}^{\,ij}(\ell) = \int \diff z\,\frac{\diff \chi}{\diff z}\,\frac{W_{X}^i(z)\,W_{Y}^j(z)}{\chi^2(z)}\,P_{XY}\left(k=\frac{\ell}{\chi(z)},z\right)\;,
\label{eq:cxy}
\end{equation}
where $X$ or $Y$ are the components that we describe in detail next, and $X, Y \in  \rm
\{\nu^e, \nu^h, \nu^g, \nu^r, g, \gamma, \gamma^I, \gamma^G\}$, corresponding to EBL, IHL, IGL, EoR, galaxy clustering, galaxy shear, intrinsic alignment, and gravitational shear, respectively. $\chi(z)$ is comoving distance along the line of sight at redshift $z$, and $W^i(z)$ is the kernel function that indicates the weight of the signal at redshift $z$ in the $i$th band or bin. The kernels determine how strongly structures at different redshifts contribute to the observed angular power spectrum, and their normalisation reflects the overall amplitude of the projected field. We show the kernels of different components as functions of redshift in Fig.\,\ref{fig:kernel}. The panels on the left show the IHL (upper) and IGL (lower) kernels of ten SPHEREx bands, respectively. The panels on the right show the kernels of the galaxy clustering (upper) and cosmic shear (lower) for six \Euclid photometric redshift bins, respectively. We  provide a detailed explanation in this section. $P_{XY}$ is the cross-power spectrum of $X$ and $Y$ at scale $k$ which is $\ell/\chi(z)$ and redshift $z$; if $X=Y$, it is an auto power spectrum. For the low-redshift components, we fix the maximum redshift $z_{\rm max}=6$, because contributions from higher redshift to the projected angular power spectra are negligible within our adopted model \citep{Cooray12,Cao20}.

\subsection{EBL}
\label{subsec:EBL}
\begin{figure}[htbp]
\centering
\includegraphics[width=0.95\columnwidth]{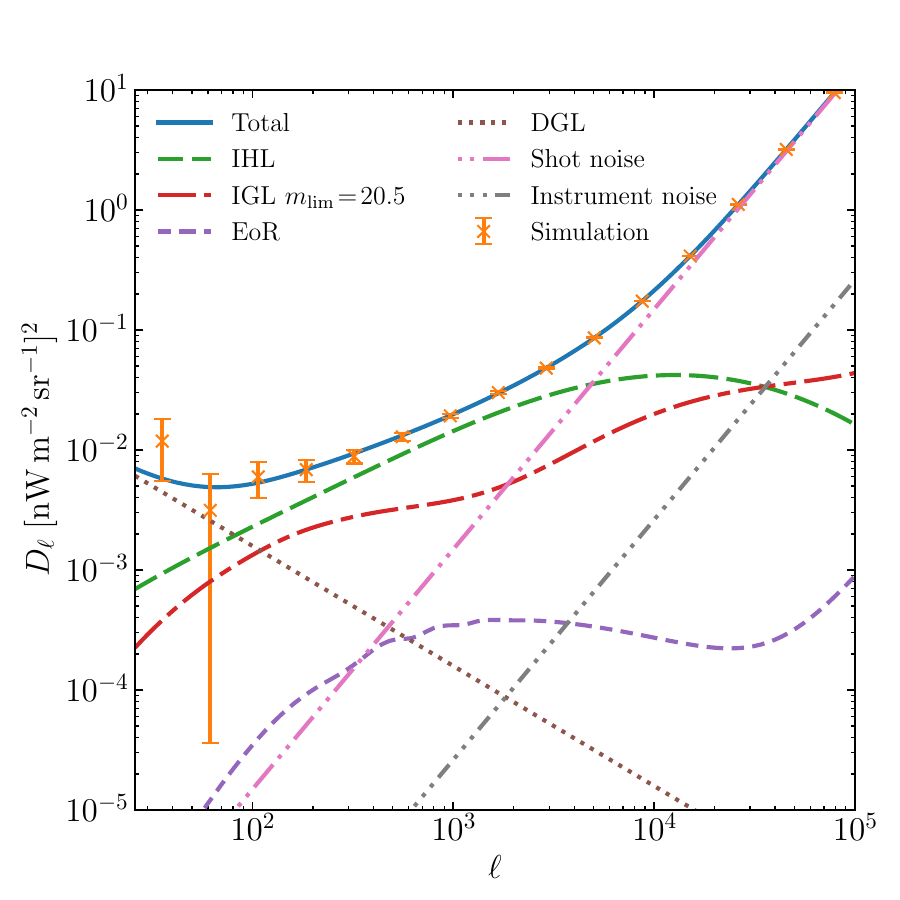}
\caption{Components of the EBL anisotropy power spectrum at 2.2\,$\si{\micron}$. Here, $D_{\ell}=\ell\,(\ell+1)\,C_{\ell}/2\pi$. The blue solid line denotes the total power spectrum with noise; the green, red, and purple dashed lines denote the power spectra of the IHL, IGL, and EoR signals, respectively. DGL is shown as a brown dotted line, for an intensity level comparable to that of the NEP deep field. The shot noise and instrument noise are shown with pink and grey dash-dotted lines, respectively. The orange crosses with error bars represent mock SPHEREx data.}
\label{fig:EBL_spectra}
\end{figure}

We use a multi-component near-infrared background model, which consists of five main components: IHL, IGL, the signal from EoR, DGL caused by the scattering of interstellar dust in the Milky Way, and shot noise from the Poisson fluctuations of unresolved sources. In Fig.\,\ref{fig:EBL_spectra}, we show the power spectra discussed above at 2.2\,$\si{\micron}$ for SPHEREx. In the following, we  discuss the detailed models of each component.

\subsubsection{IHL}
\label{subsubsec:IHL}
The IHL luminosity as a function of halo mass $M$ and redshift $z$ at the observed wavelength $\lambda$, is modelled in \cite{Cooray12} as
\begin{equation}
L_{\lambda_{\rm obs}}^{\rm IHL}(M,z)=f_{\rm IHL}(M)\,L(M,0)\,(1+z)^{\alpha}\,F_{\rm IHL}(\lambda)\;,
\label{eq:L_IHL}
\end{equation}
where $\alpha$ is the power-law index associated with redshift evolution, where we set the fiducial value $\alpha=-1.05$ \citep{Feng19}, and $f_{\rm IHL}$ is the fraction of total halo luminosity. In this work, we adopt the results from \cite{Purcell08} on the IHL fraction as a reference. We then divide $f_{\rm IHL}$ into seven halo mass bins and fit it with a power-law function, which can be written as  \citep{Cooray12}
\begin{equation}
f_{\rm IHL}(M)=A_{\rm IHL}\,\left(\frac{M}{10^{12}\,\si{\solarmass}}\right)^{\beta}\;,
\end{equation}
where $A_{\rm IHL}$ is an amplitude factor and $\beta$ is a mass power-law index. $L(M,0)$ is the total luminosity of a halo with mass $M$ at redshift $z = 0$, and its fitting function at 2.2\,$\si{\micron}$ can be represented by \citep{Lin04,Cooray12}
\begin{equation}
L(M,0)=2.76 \times 10^{12}\,h^{-2}\,\left(\frac{M}{1.9 \times 10^{14}\,h^{-1}\,\si{\solarmass} }\right)^{0.72}\,\si{\solarluminosity}\;,
\end{equation}
where $H_{0}=100\,{\rm \si{\hubble}\,km\,s^{-1}\,Mpc^{-1}}$ is the Hubble constant today, and $\si{\solarluminosity} $ is the total luminosity of the Sun. Then, we use the rest-frame spectral energy distribution (SED) $F_{\rm IHL}(\lambda)$ of the IHL to transfer $L_0$ to the other frequencies, using $\lambda=\lambda_{\rm obs}\,(1+z)^{-1}$. We assume that the SED of the IHL is described by stellar populations with an age of 3\,Gyr. 

Under these assumptions, the kernel function of IHL can be written as
\begin{equation}
 W_{{\rm \nu^h}}(z) = \frac{a^{(1-\alpha)}\,F_{\rm IHL}(\lambda)}{4\pi}\;,
\label{eq:W_IHL}
\end{equation} 
where $a$ is the scale factor. Following the halo model \citep{Cooray02}, the power spectrum of the IHL component is
\begin{equation}
{P_{{\rm \nu^h}{\rm \nu^h}}=P^{\rm 1h}_{{\rm \nu^h}{\rm \nu^h}}+P^{\rm 2h}_{{\rm \nu^h}{\rm \nu^h}}\;.}
\end{equation}
The one-halo and two-halo terms are related to small-scale fluctuations within individual haloes and the large-scale matter fluctuations, respectively. They can be represented as
\begin{equation}
\begin{aligned}
P_{{\rm \nu^h}{\rm \nu^h}}^{1h}(k) = &\int_{M_{\rm min}}^{M_{\rm max}} \diff M\,\frac{\diff n}{\diff M}\,f^2_{\rm IHL}(M)\,L^2(M,0)\,u^2(k|M)\;,\\
P_{{\rm \nu^h{\rm \nu^h}}}^{\rm 2h}(k) = &P^{\rm lin}(k)\Bigg[\int_{M_{\rm min}}^{M_{\rm max}} \diff M\frac{\diff n}{\diff M}f_{\rm IHL}(M)\,L(M,0)b(M|z)u(k|M)\Bigg]^2,
\end{aligned}
\end{equation}
where $P^{\rm lin}(k)$ is the linear matter power spectrum, $\diff n/\diff M$ is the halo mass function, and $b(M | z)$ is the linear halo bias \citep{Sheth99}, which describes how haloes with mass $M$ are biased relative to the dark matter density field, and it is accurate on large scales with $k \lesssim$ a few $\times\,0.1\,h\,{\rm Mpc^{-1}}$ where the two-halo term dominates; the correction at smaller scales is not necessary as the one-halo term dominates at higher $k$ values. The quantity $u(k|M)$ is the Fourier transform of the mass density profile of a dark matter halo with mass $M$ and redshift $z$ \citep{Navarro96}. For simplicity, we set $M_{\rm min} = 10^{10}\,\si{\solarmass} $, and $M_{\rm max} = 10^{15}\,\si{\solarmass}$ when we calculate the integral over halo mass. The haloes with masses below $M_{\rm min}$ have intrinsically low luminosities, while haloes with masses above $M_{\rm max}$ are very rare. Therefore, the contribution to the integral from outside this mass range can be neglected.

\subsubsection{IGL}
\label{subsubsec:IGL}
\begin{figure*}[htbp!]
\centering
\includegraphics[width=0.95\textwidth]{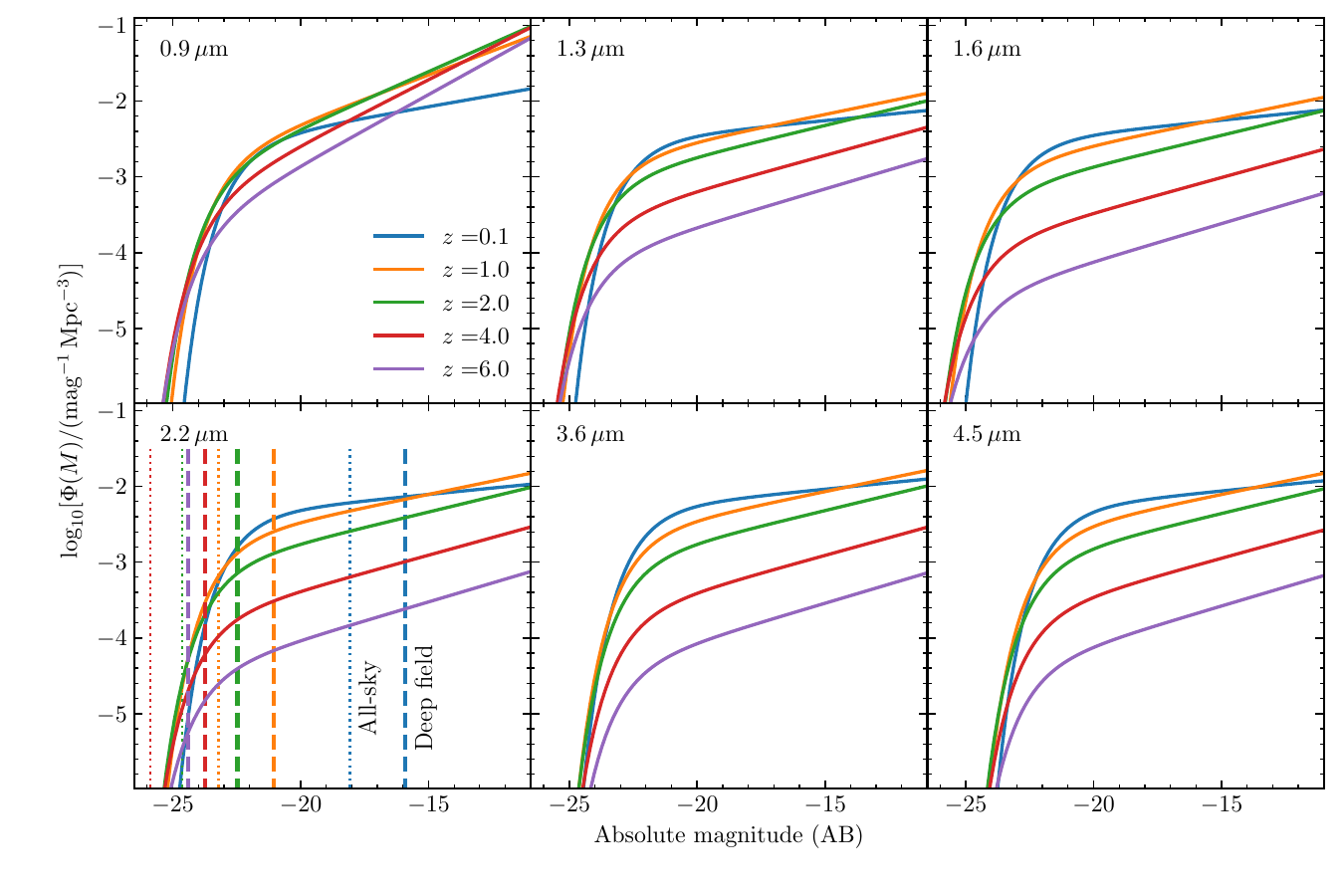}
\caption{Luminosity functions obtained using the best-fit parameters are  from \cite{Helgason12}. The dashed and dotted vertical lines are the magnitude depths of the SPHEREx deep field and all-sky survey, respectively.}
\label{fig:LF}
\end{figure*}

The luminosity function of faint galaxies is characterised by a Schechter function \citep{Schechter76}
\begin{equation}
\begin{aligned}
\Phi(M_{\rm abs} | z) =& 0.4\,\ln(10)\,\phi^*(z)\,\left\{ 10^{0.4\,\left[ M_{\rm abs}^*(z)-M_{\rm abs} \right] }\right\}^{\alpha(z)+1} \\
& \quad \times \exp\left\{ -10^{0.4\,\left[M_{\rm abs}^*(z)-M_{\rm abs}\right]}\right\}\;,
\end{aligned}
\end{equation}
where $M_{\rm abs}$ is the absolute magnitude, and we relate it to the AB magnitude by $m_{\rm AB}=M_{\rm abs} +{\rm DM}(z)-2.5\logten(1+z)$, where ${\rm DM}(z)$ is the distance modulus. The quantities $\phi^*(z)$, $M^*_{\rm abs}(z)$, and $\alpha(z)$ are the normalisation parameter, characteristic absolute magnitude, and faint-end slope at redshift $z$, respectively. \cite{Helgason12} fit the UV, optical, and NIR data and found that the individual Schechter parameters $M_{\rm abs}^*(z)$, $\phi^*(z)$, and $\alpha(z)$ can be written as
\begin{equation}
\begin{aligned}
M_{\rm abs}^*(z) &= M^*_{\rm abs,0} - 2.5\logten \left\{\left[1+(z-z_0)\right]^q\right\} \;,\\
\phi^*(z) &= \phi^*_0\,\exp\left[-p\,(z - z_0)\right] \;, \\
\alpha(z) &= \alpha_0\,(z/z_0)^r \;,
\end{aligned}
\end{equation}
where we take $z_0=0.8$ for $M_{\rm abs}^*(z)$ and $\phi^*(z)$, but $z_0= 0.01$ for $\alpha(z)$. The best-fit evolution parameters are given in Table\,\ref{tab:ep}, and the luminosity functions are shown at a range of redshifts in Fig.\,\ref{fig:LF}. We convert the magnitude depth of SPHEREx to the absolute magnitudes at different redshifts at 2.2\,$\si{\micron}$. The dashed and dotted vertical lines are the magnitude depths of deep field and all-sky survey, respectively. We then obtain the continuous evolution of the luminosity function by linearly interpolating the luminosity function number densities between adjacent wavelengths at fixed redshift.
\renewcommand{\arraystretch}{1.3}
\begin{table}
\centering
\caption{\label{tab:ep} Best-fit evolution parameters for the luminosity function.}
\begin{tabular}{ccccccc}
\hline\hline
\rule[-1mm]{0mm}{4mm}
$\lambda$ & $M_{\rm abs,0}^*$ & $q$ & $\phi^*_0$ & $p$ & $\alpha_0$ & $r$\\
$($\si{\micron}$)$ &  &  & ($\rm 10^{-3}\,Mpc^{-3}$) &  &  & \\
\hline
0.79 & $-22.80$ & 0.4 & 2.05 & 0.4 & $-1$ & 0.070\\
0.91 & $-22.86$ & 0.4 & 2.55 & 0.4 & $-1$ & 0.060\\
1.27 & $-23.04$ & 0.4 & 2.21 & 0.6 & $-1$ & 0.035\\
1.63 & $-23.41$ & 0.5 & 1.91 & 0.8 & $-1$ & 0.035\\
2.20 & $-22.97$ & 0.4 & 2.74 & 0.8 & $-1$ & 0.035\\
3.60 & $-22.40$ & 0.2 & 3.29 & 0.8 & $-1$ & 0.035\\
4.50 & $-21.84$ & 0.3 & 3.29 & 0.8 & $-1$ & 0.035\\
\hline
\end{tabular}
\tablefoot{These values are  from \cite{Helgason12}.}
\end{table}

The galaxy number counts in each magnitude bin per unit solid angle can be obtained through
\begin{equation}
N(m) = \int \Phi(M_{\rm abs} | z)\, \chi^2(z) \, \frac{\diff \chi}{\diff z} \, \diff z\;,
\end{equation}
the flux production rate is given by
\begin{equation}
\frac{\diff \mathcal{F}}{\diff z}=\int_{m_{\rm lim}}^{\infty} \diff m\,f(m)\, \frac{\diff N(m|z)}{\diff z}\;,
\end{equation}
where $m_{\rm lim} = 20.5$ at 2.2\,$\si{\micron}$ is the masking threshold \citep{Cheng21}. Brighter galaxies in EBL measurements are predominantly foreground objects and therefore need to be masked. This is a simplifying choice adopted for the forecast, and it does not fully capture possible wavelength-dependent variations in the masking depth across bands. Although the total EBL intensity is given by the integral of all background galaxies, bright background galaxies are rare and contribute significantly only locally, with a negligible effect on the angular power spectrum. We mask sources brighter than the masking threshold, resulting in the removal of approximately 50\% of the pixels. Here $f(m)=\nu f_{\nu}(m)$, and the specific flux $f_{\nu}(m) = 10^{-0.4(m-23.9)}\,{\rm \mu Jy}$, and $\mathcal{F}$ is the integrated flux in units of $\rm nW\,m^{-2}\,sr^{-1}$. Thus the kernel function of the IGL can be written as
\begin{equation}
W_{{\rm \nu^g}}(z) =\frac{\diff z}{\diff \chi}\,\frac{\diff \mathcal{F}}{\diff z}\;.
\label{eq:W_IGL}
\end{equation}

The total, one-halo and two-halo terms of the power spectrum for the IGL are given by
\begin{equation}
P_{{\rm \nu^g}{\rm \nu^g}}=P_{\rm gg}^{\rm 1h}(k) +P_{\rm gg}^{\rm 2h}(k)\;,
\end{equation}
and
\begin{equation}
\begin{aligned}
P_{\rm gg}^{\rm 1h}(k) & =\int \diff M\,\frac{\diff n}{\diff M}\,\frac{2\left\langle N_{\rm sat}\right\rangle\,\left\langle N_{\rm  cen}\right\rangle\, u(k | M)+\left\langle N_{\rm sat}\right\rangle^2\,u^2(k | M)}{\bar{n}_{\rm gal}^2}\;,\\
P_{\rm gg}^{\rm 2h}(k) & =P^{\rm lin}(k)\,\left[\int \diff M\,\frac{\diff n}{\diff M}\,\frac{\langle N_{\rm gal}\rangle}{\bar{n}_{\rm gal}}\,b(M)\,u(k|M)\right]^2\;,
\end{aligned}
\end{equation}
where $\bar{n}_{\rm gal}$ is the average number density of galaxies 
\begin{equation}
\bar{n}_{\rm gal}=\int \diff m\,\frac{\diff n}{\diff M}\,\langle N_{\rm gal} \rangle\;.
\end{equation}
The number of galaxies per halo is divided into central galaxies, $\langle N_{\rm cen}\rangle$, and satellite galaxies, $\langle N_{\rm sat}\rangle$, such that $\langle N_{\rm gal}\rangle = \langle N_{\rm sat}\rangle + \langle N_{\rm cen}\rangle$. In the halo occupation distribution (HOD) model, the expected number can be expressed as
\begin{equation}
\begin{aligned}
\langle N_{\rm cen}\rangle & = \frac{1}{2} \left\{ 1 + {\rm erf} \left[ \frac{\logten (M/M^*_{\rm min})}{\sigma_{M}} \right]\right\}\;; \\
\langle N_{\rm sat}\rangle & = \frac{1}{2} \left\{ 1 + {\rm erf} \left[ \frac{\logten(M/2M^*_{\rm min})}{\sigma_{M}} \right]\right\}\, \left( \frac{M}{M_{\rm s}} \right)^{\alpha_{\rm s}}\;.
\end{aligned}
\end{equation}
Here, $\sigma_{M}$ is the dispersion in halo mass, which controls the smoothness of the transition, and $M^*_{\rm min}$ is the characteristic minimum halo mass for hosting a central galaxy. The cutoff mass for the satellite term is twice that of the central galaxy and grows as a power law with a slope of $\alpha_{\rm s}$, normalised by $M_{\rm s}$ \citep{Zheng05}. We adopt the HOD model parameters motivated by SDSS measurements, and set $\sigma_{M} = 0.2$, $M^*_{\rm min} =10^9\,\si{\solarmass}$, $\alpha_{\rm s}=1$, and $M_{\rm s} =5 \times 10^{10}\,\si{\solarmass}$ \citep{Helgason12,Thacker15}. Although fixing certain parameters necessarily imposes modelling assumptions, these parameters are weakly constrained by the data or subject to strong priors, and fixing them therefore improves the fitting accuracy and reduces the computational cost.

\subsubsection{EoR}
\label{subsubsec:EoR}
\begin{figure*}[htbp!]
    \centering
    \includegraphics[width=0.49\textwidth]{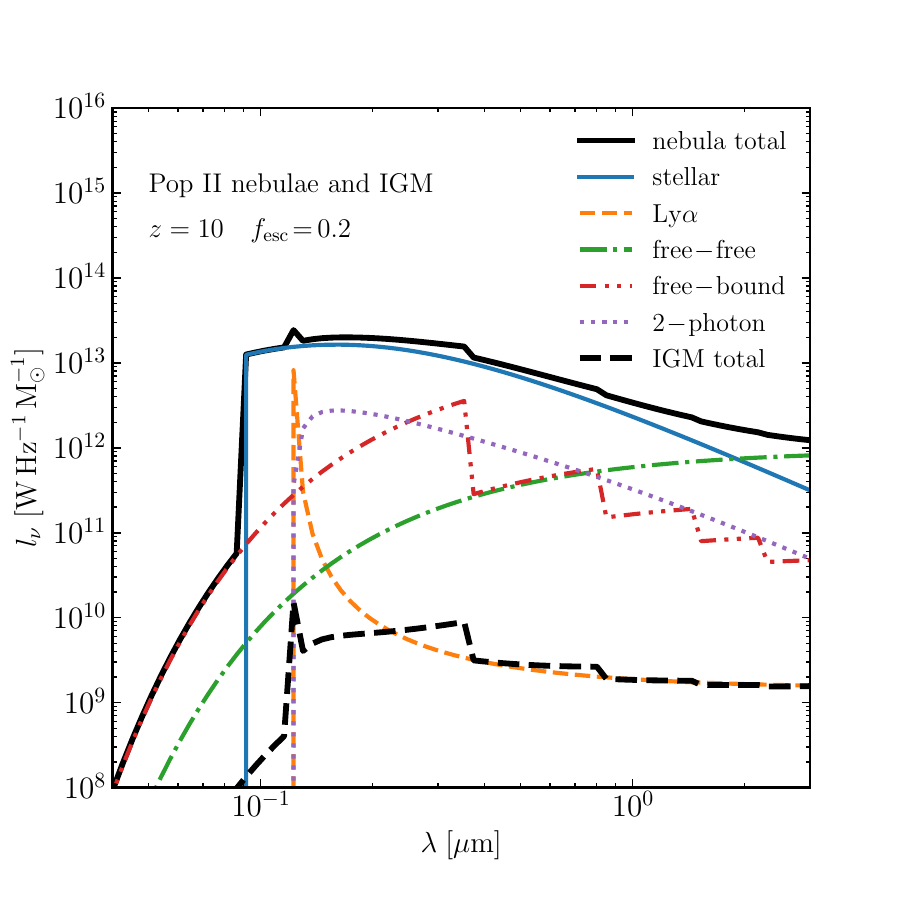}
    \includegraphics[width=0.49\textwidth]{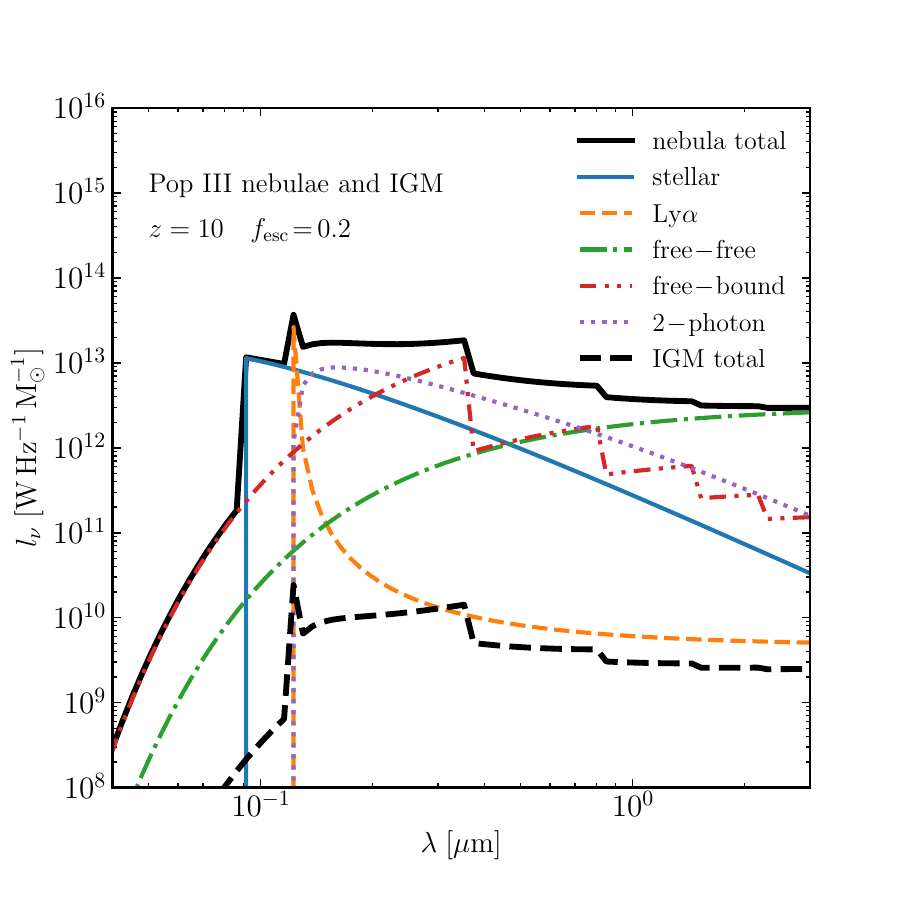}
    \caption{Luminosity mass density $l_\nu$ as a function of the rest-frame wavelength $\lambda$ for Pop II (\emph{Left}) and Pop III (\emph{Right}) stars at $z = 10$. The black solid and dashed lines are the total $l_\nu$ from the stellar nebulae and the IGM. The coloured lines are the various components of stellar nebulae emission: the direct emission from the stars; the $\rm{Ly}\alpha$ line; and free-free, free-bound, and two-photon processes. We set $f_{\rm esc}$ = 0.2 for the Pop II and the Pop III cases.}
    \label{fig:lv}
\end{figure*}

To model the signal from EoR, we model the emission from the Pop III and early Pop II stars \citep{Cooray12b,Sun21}. The power spectrum of EoR is given by $P_{{\rm \nu^r}{\rm \nu^r}} = P_{gg}^{\rm 1h}(k) +P_{\rm gg}^{\rm 2h}(k)$, and the kernel function of EoR can be written as
\begin{equation}
 W_{{\rm \nu^r}}(z) = a\,\bar{\,j}_{\nu}(z)\;,
\label{eq:W_EoR}
\end{equation}
where $\bar{\,j}_\nu(z)$ is the mean emission per comoving volume at frequency $\nu=(z+1)\,\nu_{\rm obs}$ and redshift $z$, the emission comes from Pop III and early Pop II stars during the reionisation period, and $\bar{\,j}_\nu(z)$ can be written by
\begin{equation}
\bar{\,j}_\nu(z)=f_{\rm P}\,\bar{\,j}_\nu^{\,\rm PopIII}(z)+\left(1-f_{\,\rm P}\right)\,\bar{\,j}_\nu^{\,\rm PopII}(z)\;.
\end{equation}
Here, $f_{\rm{P}}$ is the population fraction, which as a redshift dependent fraction can be expressed as
\begin{equation}
f_{\rm P}(z)=\frac{1}{2}\left[1+\operatorname{erf}\left(\frac{z-10}{\sigma_{\rm P}}\right)\right]\;,
\end{equation}
where $\sigma_{\rm P} = 0.5$ is the population transition width. The quantities $\bar{\,j}_\nu^{\,\rm PopII}$ and $\bar{\,j}_\nu^{\,\rm PopIII}$ are the comoving specific emission coefficients, and can be written as
\begin{equation}
\bar{\,j}_\nu^{\,\rm Pop}(z)=\frac{1}{4 \pi}\, l_\nu^{\,\rm Pop}\,\left\langle\tau_*^{\,\rm Pop}\right\rangle\, \psi(z)\;,
\end{equation}
where $\left\langle\tau_*\right\rangle$ is the mean lifetime of each stellar type, adopted following \cite{Cooray12}. The quantity $l_\nu$ is the luminosity density at frequency $\nu$, the total $l_\nu$ emitted from the stellar nebulae and scattered in the intergalactic medium (IGM; \citealt{Santos02,Cooray12b}), which includes several components, such as the direct emission from the stellar $l_\nu^*$, as well as the $\rm{Ly}\alpha$ line $l_\nu^{\,\rm Ly\alpha}$, free-free $l_\nu^{\,\rm ff}$, free-bound $l_\nu^{\,\rm fb}$, and two-photon processes $l_\nu^{\,\rm 2ph}$. We can write the luminosity mass density as
\begin{equation}
\begin{aligned}
l_\nu^{\,\rm neb}&=l_\nu^*+\left(1-f_{\rm esc}\right)\,\left(l_\nu^{\,\rm Ly\alpha,neb}+l_\nu^{\,\rm ff,neb}+l_\nu^{\,\rm fb,neb}+l_\nu^{\,\rm 2ph,neb}\right)\;,\\
l_\nu^{\,\rm IGM}&=f_{\rm esc}\,\left(l_\nu^{\,\rm Ly\alpha,IGM}+l_\nu^{\,\rm ff,IGM}+l_\nu^{\,\rm fb,IGM}+l_\nu^{\,\rm 2ph,IGM}\right)\;.
\end{aligned}
\end{equation}
We show the luminosity mass density $l_\nu$ at rest-frame wavelength for Pop II and Pop III stars in Fig.\,\ref{fig:lv}. The black solid and dashed lines are the total $l_\nu$ from the stellar nebulae and the IGM. The coloured lines are the various components of stellar nebulae emission. For both Pop II and Pop III cases, we set $f_{\rm esc}$ = 0.2 \citep{Bauer15,Ma20}. Raising $f_{\rm esc}$ would lead to weaker Ly$\alpha$ signal. We note that due to the absorption by neutral hydrogen gas, we cut off the portion with wavelengths shorter than Ly$\alpha$.
 
For the comoving star formation rate density (SFRD) $\psi(z)$, we adopt ongoing star formation model \citep{Santos02}, which is given by
\begin{equation}
\psi(z)=f_*(z)\,\frac{\Omega_{\rm b}}{\Omega_{\rm m}}\,\frac{\diff}{\diff t} \int_{M_{\rm \min}}^{\infty} \diff M\, M\,\frac{\diff n}{\diff M}(M, z)\;,
\end{equation}
where $f_*(z)$ is the star formation efficiency, which denotes the fraction of baryons converted to stars at redshift $z$. We used the total SFRD form \cite{Finkelstein16} to fit $\psi(z)$ at $4 < z \le10$, and set $f_*(z)=0.01$ at $ z > 10$. The mid-point reionisation redshift is $z_{\rm re} =7.7$ \citep{Aghanim20}, and we take the range from $z_{\rm min}= 6 $ to $z_{\rm max}=30$, as the redshifts of the end and beginning of EoR.

\subsubsection{DGL}
\label{subsubsec:DGL}
DGL refers to the light that is emitted by stars and scattered by interstellar dust within a galaxy, and it dominates the contribution to the EBL at certain frequencies on large scales \citep{Brandt12}. Since it is proportional to the size and density of the dust, its angular power spectrum is related to the emission from interstellar dust. At high Galactic latitudes, the power spectrum of DGL approximately follows a simple power law, as $C_{\ell}\propto \ell^{-3}$ \citep{Zemcov14}, and we  consider a model parameter $A_{\rm DGL}$ as the amplitude factor of a uniform interstellar radiation field. Following \cite{Sano17}, we model the DGL as the sum of a scattered light component and a dust thermal emission component and write it as
$I_\lambda^{\rm{DGL}}=I_{\lambda, \rm{sca}}+I_{\lambda, \rm{em}}$,
where the scattering light component $I_{\lambda, \rm{sca}}$ is created using the interstellar dust model developed by \citep{Weingartner01}, and the thermal emission component $I_{\lambda, \rm{em}}$ is described using the dust emission model of \citep{Draine07}. The DGL in each band using the $100\,\si{\micron}$ dust map as a spatial template, and the DGL intensity in $i$-band is assumed to scale with the $100\,\si{\micron}$ emission as $I_{\lambda_i}^{\rm DGL}(b)=R_i(b)\,I_{100}(b)$, where $R_i(b)$ is the Galactic latitude dependent conversion factor at wavelength $\lambda_i$. For the cross-correlation between the $i$th and $j$th bands, we model the DGL amplitude as $A_{\rm DGL}^{ij} = R_i(b)R_j(b)A_{100}(b)$, where $A_{100}(b)$ is the angular power spectrum amplitude of the $100\,\si{\micron}$ dust template.

\subsubsection{Shot noise}
\label{subsubsec:Shot}
Shot noise is crucial in fluctuation measurements as it helps identify unresolved galaxy populations \citep{Kashlinsky07}, and reflects the statistical counting noise of unresolved sources within the instrument beam. At small scales, shot noise dominates the contribution of non-linear clustering of galaxies to the power spectrum, such that the observed flat power spectrum is entirely attributed to shot noise. We assume that the shot-noise power spectrum is simply $P_{\rm SN} = {\rm const}$ (observed as $P_{\rm SN}\times|B_{\ell}P_{\ell}|^2$ after beam and pixel convolution), which can be expressed as
\begin{equation}
P_{\rm SN} = \int \diff z \int_{m_{\rm lim}}^{\infty} \diff m\,f^2(m)\,\frac{\diff N(m|z)}{\diff z}\;,
\end{equation}
where $m_{\rm lim}$ is used as the magnitude threshold to separate resolved and/or removed galaxies from the unresolved remaining sources. This model allows us to evaluate the effect of a given limiting magnitude $m_{\rm lim}$ on the shot noise level, and calculate the shot noise associated with unresolved galaxies. 

\subsection{Cosmic shear}
\label{subsec:Shear}
Considering the impact of residual additive and multiplicative systematic effects \citep{Amara08}, the angular power spectrum of observed galaxy shear for the $i$th and $j$th tomographic redshift bins is given by
\begin{equation}
{\hat{C}^{\,ij}_{\gamma\gamma}(\ell)=(1+m_i)\,(1+m_j)\, C_{\gamma\gamma}^{\,ij}(\ell) +N^{ij}_{\rm add}(\ell) + \frac{\delta_{ij}\,\sigma_\gamma^2}{\bar{n}_i}\;.}
\end{equation}
Due to the dependence of the multiplicative systematic term on the signal, we introduce a calibration parameter $m_i$ in the $i$-th tomographic redshift bin to relate the signal power spectrum {$C_{\gamma\gamma}^{\,ij}(\ell)$}to the observed power spectrum {$\hat{C}_{\gamma\gamma}^{\,ij}(\ell)$}. Additive shear bias $N^{ij}_{\rm add}$ is typically corrected at the catalogue level and is expected to be uncorrelated with the true shear signal, therefore it only contributes to noise and is neglected $N^{ij}_{\rm add}=0$ \citep{Asgari21,Amon22}. Here $\sigma_\gamma = 0.3$ is the intrinsic ellipticity variance, which is taken from \cite{Blanchard-EP7}, and $\bar{n}_i$ is the average galaxy number density of galaxies per steradian in the $i$-th tomographic redshift bin.

\begin{figure}[htbp]
\centering
\includegraphics[width=0.95\columnwidth]{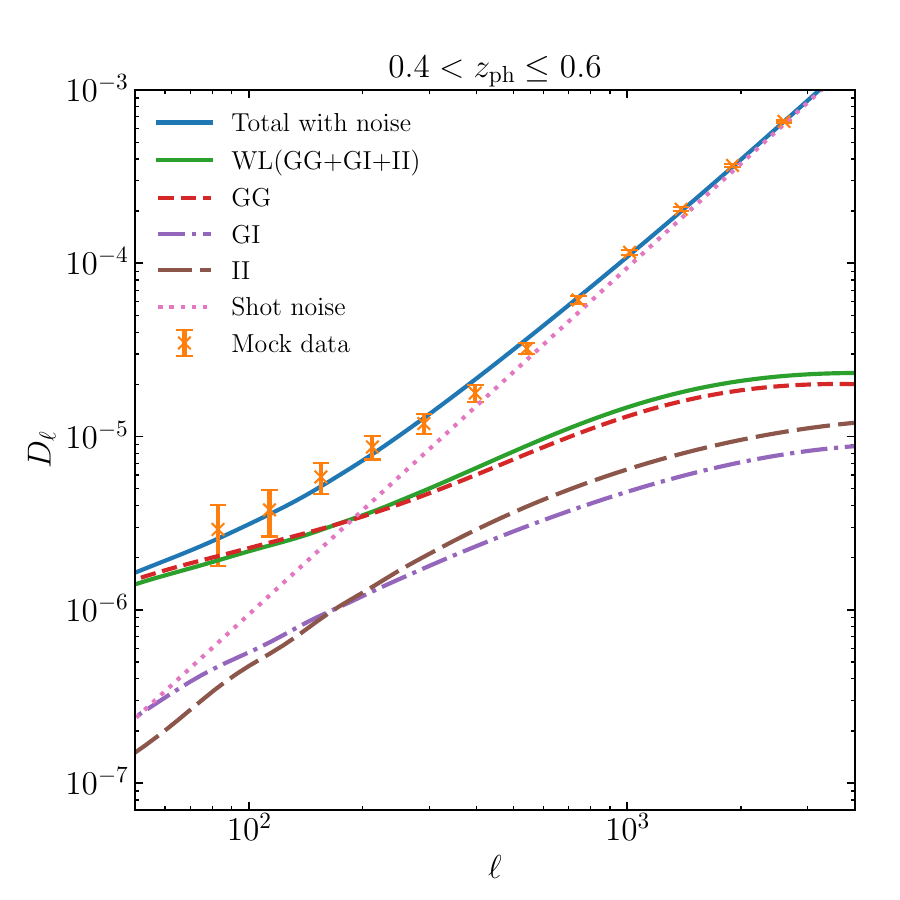}
\caption{ Components of the cosmic shear power spectrum for the tomographic redshift bin with $0.4\!<\!z_{\rm ph}\!\le\!0.6$. The blue and green solid lines denote the total power spectrum with noise and the power spectrum of the signal, respectively. The additive shot noise is shown as a pink dotted line. The red dashed, brown dashed, and purple dot-dashed curves are the power spectra of gravitational-gravitational, intrinsic-intrinsic, and gravitational-intrinsic power spectra, respectively. We note that the value of $C_{{\rm \gamma^G}{\rm \gamma^I}}(\ell)$ is negative. The orange crosses with error bars represent mock \Euclid cosmic shear mock data.}
\label{fig:WL_spectra}
\end{figure}

When inferring the weak gravitational lensing between galaxies and the observer through galaxy shear measurements, the results are impacted by IA \citep{Troxel15}. The observed shape signal is composed of two components, namely the gravitational shear and intrinsic shape, expressed as $\gamma^{\rm obs} = \gamma^{\rm G} + \gamma^{\rm I}$, and the total shear power spectrum is given by \citep{Hirata04,Bridle07,Secco22,Cao24}
\begin{equation}
{C_{\gamma\gamma}^{\,ij}(\ell)=C_{{\rm \gamma^G}{\rm \gamma^G}}^{\,ij}(\ell)+C_{{\rm \gamma^G}{\rm \gamma^I}}^{\,ij}(\ell)+C_{{\rm \gamma^I}{\rm \gamma^I}}^{\,ij}(\ell)\;,}
\end{equation}
where $C_{{\rm \gamma^G}{\rm \gamma^G}}^{\,ij}(\ell)$ is the convergence power spectrum, which is obtained form the standard lensing signal, and $C_{{\rm \gamma^G}{\rm \gamma^I}}^{\,ij}(\ell)$ and $C_{\rm{{\rm \gamma^I}{\rm \gamma^I}}}^{\,ij}(\ell)$ are the gravitational-intrinsic (GI) and intrinsic-intrinsic (II) power spectra, respectively. GI denotes the correlation of the gravitational shear of a foreground or background galaxy and the intrinsic shape of a background or foreground galaxy between different bins, and II denotes the correlation of the intrinsic shapes. Under the Limber approximation \citep{Limber53,LoVerde08}, the above spectra can be written as
\begin{equation}
\begin{aligned}
{C}^{\,ij}_{{\rm \gamma^G}{\rm \gamma^G}}\left(\ell\right) =& \int \diff z\,\frac{\diff \chi}{\diff z}\,\frac{W_{{\rm \gamma^G}}^{\,i}(z)\,W_{{\rm \gamma^G}}^{\,j}(z)}{\chi^2(z)}\,P_{{\rm \gamma^G}{\rm \gamma^G}}\left(k=\frac{\ell}{\chi(z)},z\right)\;,\\
{C}^{\,ij}_{{\rm \gamma^I}{\rm \gamma^I}}\left(\ell\right) =& \int \diff z\,\frac{\diff \chi}{\diff z}\,\frac{W_{{\rm \gamma^I}}^{\,i}(z)\,W_{{\rm \gamma^I}}^{\,j}(z)}{\chi^2(z)}\,P_{{\rm \gamma^I}{\rm \gamma^I}}\left(k=\frac{\ell}{\chi(z)},z\right)\;,\\
{C}^{\,ij}_{{\rm \gamma^G}{\rm \gamma^I}}\left(\ell\right) =& \int \diff z\,\frac{\diff \chi}{\diff z}\,\frac{W_{{\rm \gamma^G}}^{\,i}(z)\,W_{{\rm \gamma^I}}^{\,j}(z)}{\chi^2(z)}\,P_{{\rm \gamma^G}{\rm \gamma^I}}\left(k=\frac{\ell}{\chi(z)},z\right)\;,\\
&+ \int \diff z\,\frac{\diff \chi}{\diff z}\,\frac{W_{{\rm \gamma^I}}^{\,i}(z)\,W_{{\rm \gamma^G}}^{\,j}(z)}{\chi^2(z)}\,P_{{\rm \gamma^G}{\rm \gamma^I}}\left(k=\frac{\ell}{\chi(z)},z\right)\;.
\end{aligned}
\end{equation}
Here, $W_{{\rm \gamma^I}}^i\left(z \right)$ is equal to the normalised distribution $n^i\left(z \right)$ in the $i$th tomographic redshift bin, and $W_{{\rm \gamma^G}}(z)$, the lensing weighting function, can be given by
\begin{equation}
W_{{\rm \gamma^G}}^i(z) =\frac{3\Omega_{\rm m}\,H_0^2}{2c^2}\,\frac{\chi(z)}{a}\, \int_z^{z_{\rm H}} \diff z'\,n^i\left(z' \right)\,\frac{\chi(z')-\chi(z)}{\chi(z')}\;,
\end{equation}
where $c$ is the speed of light, $z_{\rm H}$ is the maximum redshift of the source distribution, and $n^i(z)$ denotes the redshift distribution of source galaxies, which is normalised such that $\int n^i(z)\,\diff z=1$. $P_{{\rm \gamma^G}{\rm \gamma^G}}$ is equal to $P_{\delta\delta}$, which is the matter power spectrum, which we calculate using the Halofit model \citep{Takahashi12}. We adopt the TATT model \citep{Blazek19} to predict $P_{{\rm \gamma^I}{\rm \gamma^I}}$ and $P_{{\rm \gamma^G}{\rm \gamma^I}}$, and they can be expressed by
\begin{equation}
\begin{aligned}
P_{{\rm \gamma^G}{\rm \gamma^I}}\left(k\right)=&\ C_1  P_\delta(k)+C_{1 \delta}\left[A_{0 \mid 0 E}\left(k\right)+C_{0 \mid 0 E}\left(k\right)\right]\\
&+ C_2\left[A_{0 \mid E 2}\left(k\right)+B_{0 \mid E 2}\left(k\right)\right]\;,\\
P_{{\rm \gamma^I}{\rm \gamma^I}}\left(k\right)=&\ C_1^2  P_\delta(k)+2 C_1 C_{1 \delta} \left[A_{0 \mid 0 E}\left(k\right)+C_{0 \mid 0 E}\left(k\right)\right]\\
&+ C_{1 \delta}^2 A_{0 E \mid 0 E}\left(k\right)+C_2^2 A_{E 2 \mid E 2}\left(k\right)\\
&+2 C_1 C_2 \left[A_{0 \mid E 2}\left(k\right)+B_{0 \mid E 2}\left(k\right)\right]\\
&+2 C_{1 \delta} C_2 D_{0 E \mid E 2}\left(k\right)\;.
\end{aligned}
\end{equation}
Here, $k$-dependent terms are discussed in \cite{Blazek19}, we calculate them using the FAST-PT module of Core Cosmology Library(CCL) \citep{Chisari19}. Under linear galaxy biasing, $C_{1 \delta}$ is generally given by $C_{1 \delta} = b_1 C_1$.
Following the DES Year 1 tomographic cosmic shear analysis \citep{Troxel18}, we fix $b_1=1$. $C_1$ and $C_2$ can be written as
\begin{equation}
\begin{aligned}
C_1=& -A_{\rm IA_1} \bar{C}_1 \rho_{\text {crit}} \frac{\Omega_{\mathrm{m}}}{D(z)}\;,\\
C_2=& 5 A_{\rm IA_2} \bar{C}_1 \rho_{\text {crit}} \frac{\Omega_{\mathrm{m}}}{D^2(z)}\;.
\end{aligned}
\end{equation}
Here, the empirical amplitude $\bar{C}_1= 5 \times 10^{-14}\, h^{-2}\si{\solarmass}^{-1}\rm{Mpc^3}$, taken from \cite{Brown02}, $\rho_{\rm{crit}}$ is the present critical density, ${D(z)}$ is the linear growth factor, normalised to unity at $z=0$, and $A_{\rm IA_1}$ and $A_{\rm IA_2}$ are two free parameters. Following \citep{Chisari19}, we set the fiducial values of $A_{\rm IA_1}$ and $A_{\rm IA_2}$ to unity in the following forecasts. As an example, in Fig.\,\ref{fig:WL_spectra}, we show the power spectra of cosmic shear for the tomographic redshift bin with $0.4\!<\!z_{\rm ph}\!\le\!0.6$. 

\subsection{Galaxy clustering}
\label{subsec:Gal}
Considering the impact of the shot-noise term, the total angular power spectrum of galaxy clustering between the $i$th and $j$th redshift bins can be written as
\begin{equation}
\hat{C}^{\,ij}_{{\rm g}{\rm g}}(\ell)= C_{{\rm g}{\rm g}}^{\,ij}(\ell)+ \frac{\delta_{ij}}{\bar{n}_i}\;,
\end{equation}
where $\bar{n}_i$ is the average galaxy number density in the $i$th redshift bin per steradian. The angular power spectrum of galaxy clustering can be calculated using Eq.\,(\ref{eq:cxy}) with the redshift kernel $W^i_{{\rm g}} = n^i(z)$ and the power spectrum of galaxy clustering is $P_{{\rm g}{\rm g}} = P_{\rm gg}^{\rm 1h}(k) +P_{\rm gg}^{\rm 2h}(k)$. In photometric surveys, although there is some overlap in the true redshift distributions of different photometric redshift bins, the proportion of galaxies in the overlapping regions is quite small, especially for two non-adjacent photometric redshift bins. In this case, the cross-power spectrum signal is very weak, and in this work we only consider the auto-power spectrum in galaxy clustering.

\subsection{Cross-correlation of EBL and cosmic shear}
\label{subsec:EBLxWL}
We also need to consider the multiplicative shear calibration and IA in the cross-correlation analysis. Therefore, the cross angular power spectrum between EBL in the $i$th band and cosmic shear in the $j$th tomographic redshift bin can be written
\begin{equation}
\begin{aligned}
{\hat{C}_{{\rm \nu^e}\gamma}^{\,ij}(\ell)}=\left(1+m_j\right)\,\Big[C_{{\rm \nu^h}{\rm \gamma^G}}^{\,ij}(\ell)+C_{{\rm \nu^g}{\rm \gamma^G}}^{\,ij}(\ell)+C_{{\rm \nu^h}{\rm \gamma^I}}^{\,ij}(\ell)\\
+C_{{\rm \nu^g}{\rm \gamma^I}}^{\,ij}(\ell) \Big]+P_{\rm{SN_x}}^{\,ij}\;,
\end{aligned}
\label{eq:EBLxWL}
\end{equation}
where $P_{\rm SN_{x}}^{\,ij}$ is the cross-shot noise between EBL and cosmic shear, which can be expressed as
\begin{equation}
P_{\rm SN_{x}}^{\,ij} = \frac{\sigma_\gamma}{\bar{n}_j}\int \diff z \int_{m_{\rm{lim}}}^{\infty} \diff m\, f_i(m)\, \frac{\diff N(m|z)}{\diff z}\;.
\end{equation}

The non-linear density $\delta$ can be expanded in terms of the linear density field and relevant gravity kernels in standard perturbation theory; it can be written as
\begin{equation}
\delta=\delta^{(1)}+\delta^{(2)}+\delta^{(3)}+\cdots\;,
\end{equation}
and we assume that the IHL field can be expressed as 
\begin{equation}
{\rm \nu^h}=C^{\rm h}_1\delta^{(1)}+C^{\rm h}_2\delta^{(2)}+C^{\rm h}_3\delta^{(3)}+\cdots\;.
\end{equation}
We express the intrinsic galaxy shape field up to second order in the linear density field as
\begin{equation}
\gamma_{i j}^I(\mathbf{x})=C_1 s_{i j}+C_2\left(s_{i k} s_{k j}\right)+C_{1 \delta}\left(\delta s_{i j}\right)+\cdots\;,
\end{equation}
where $s_{i j}$ is the gravitational tidal field, which at any given position $\mathbf{x}$ is a $3\times3$ tensor. By analogy with $P_{{\rm \gamma^G \gamma^G}}$ and $P_{{\rm \gamma^G \gamma^I}}$, the corresponding cross-spectrum can be written as
\begin{equation}
\begin{aligned}
P_{\rm \nu^h \gamma^G}\left(k\right)=&\ C^{\rm h}_1 P_{\delta^{(1)}}(k) +C^{\rm h}_2 P_{\delta^{(2)}}(k)\\
P_{\rm \nu^h \gamma^I}\left(k\right)=&\ C_1 \left[C^{\rm h}_1 P_{\delta^{(1)}}(k) +C^{\rm h}_2 P_{\delta^{(2)}}(k)\right]\\
&+C_{1 \delta}\left[C^{\rm h}_2 A_{0 \mid 0 E}\left(k\right)+C^{\rm h}_1 C_{0 \mid 0 E}\left(k\right)\right]\\
&+ C_2\left[C^{\rm h}_2 A_{0 \mid E 2}\left(k\right)+ C^{\rm h}_1 B_{0 \mid E 2}\left(k\right)\right]\;.
\end{aligned}
\end{equation}

\begin{figure*}[htbp!]
\center 
\includegraphics[width=0.48\textwidth]{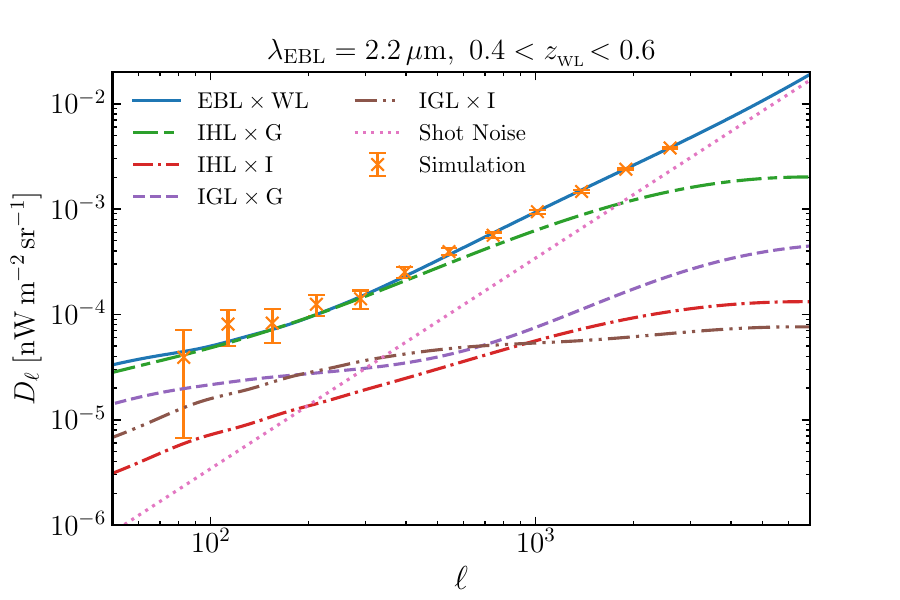}
\includegraphics[width=0.48\textwidth]{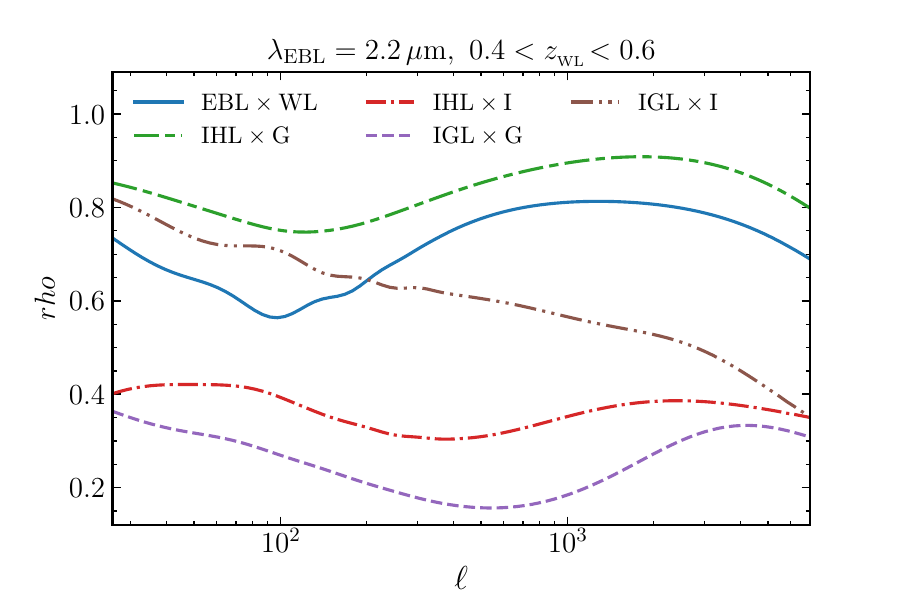}

\includegraphics[width=0.48\textwidth]{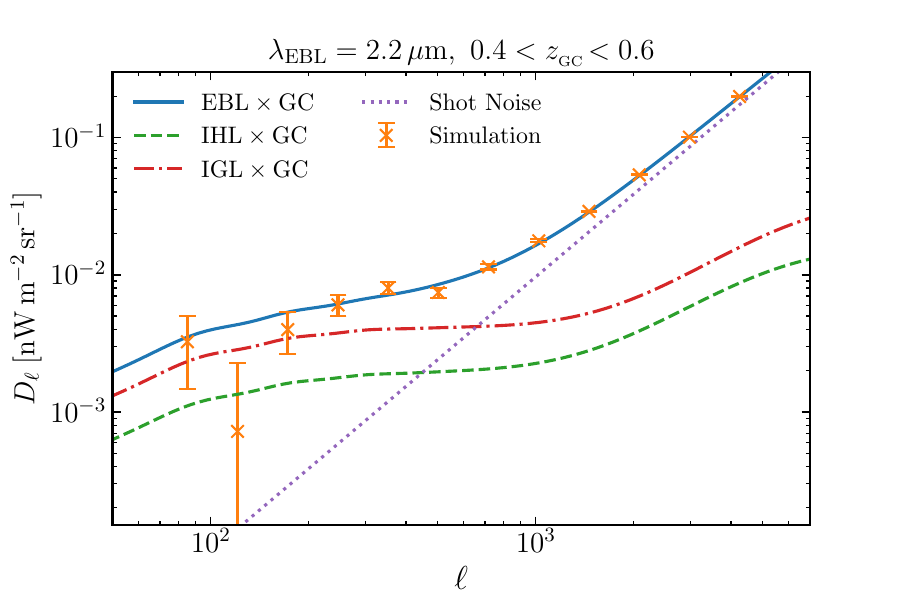} 
\includegraphics[width=0.48\textwidth]{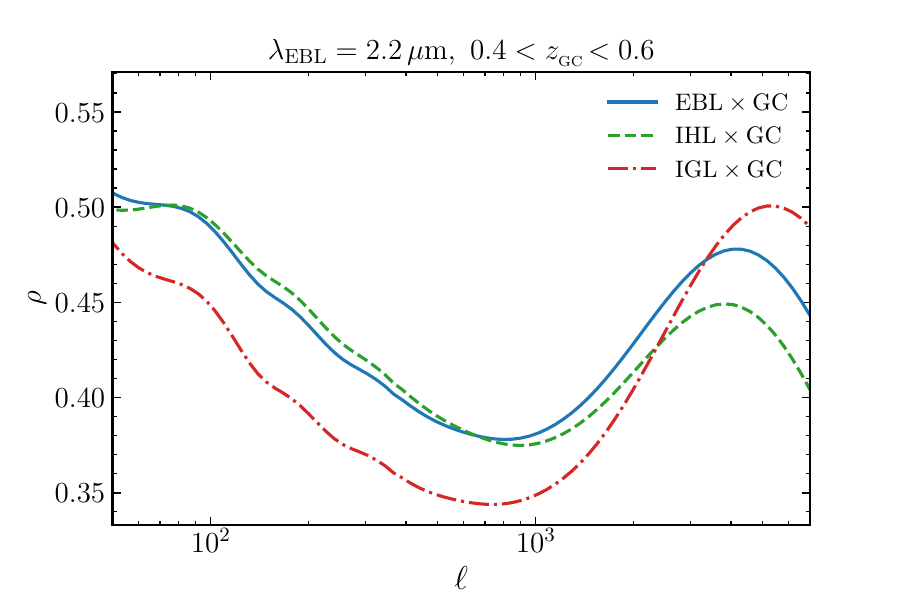}
\caption{Components of the cross angular power spectrum between EBL at 2.2\,$\si{\micron}$ and cosmic shear--galaxy clustering at the tomographic redshift bin with $0.4\!<\!z_{\rm ph}\!<\!0.6$ (left panels) and their correlation coefficient (right panels). We use the same colour and type of line to present the same component in the left and right panels. \emph{Top}: Solid blue  lines denote the cross-correlation between the total components of EBL, and the green, red, purple, and brown lines represent the cross-correlation of IHL-gravitational (IHL$\times$G), IHL-intrinsic (IHL$\times$I), IGL-gravitational (IGL$\times$G), and IGL-intrinsic (IGL$\times$I) signals, respectively. We note that the value of IHL$\times$I and IGL$\times$I are negative. \emph{Bottom}: Solid blue lines represent the cross-correlation of the total EBL with galaxy clustering. The green and red lines are the IHL-galaxy clustering (IHL$\times$GC) and IGL-galaxy clustering (IGL$\times$GC) signals, respectively.}
\label{fig:cross_sp}
\end{figure*}

We present the cross-correlation power spectra and correlation coefficient between EBL at 2.2\,$\si{\micron}$ and cosmic shear in the tomographic photometric redshift bin of $0.4\!<\!z\!<\!0.6$ as an example in Fig.\,\ref{fig:cross_sp}. We find that the correlation coefficient between the IGL and cosmic shear is smaller than that between the IHL and cosmic shear, especially for IGL$\times$G, which is over 50\% lower than IHL$\times$G. This suggests that cross-correlation can be effectively utilised to separate the IHL and IGL components.

\subsection{Cross-correlation of EBL and galaxy clustering}
\label{subsec:EBLxGC}
In this section, we discuss the angular cross-power spectra of EBL and galaxy clustering in our model. The total angular cross-power spectrum is composed of two components, which are
\begin{equation}
C_{{\rm \nu^e}{\rm g}}^{\,ij}(\ell)=C_{{\rm \nu^h}{\rm g}}^{\,ij} (\ell)+C_{{\rm \nu^g}{\rm g}}^{\,ij}(\ell)+P_{\rm SN_{xx}}^{\,ij}\;,
\end{equation}
where $P_{\rm SN_{xx}}^{\,ij}$ is the EBL-galaxy clustering cross-shot noise, which
can be expressed as
\begin{equation}
P_{\rm SN_{xx}}^{\,ij} = \frac{1}{\bar{n}_j}\int \diff z \int_{m_{\rm{lim }}}^{\infty} \diff m\, f_i(m)\,\frac{\diff N(m|z)}{\diff z}\;,
\end{equation}
while $C_{\rm{{\rm \nu^h}{\rm g}}}^{\,ij}(\ell)$ and $C_{{\rm \nu^g}{\rm g}}^{\,ij}(\ell)$ are cross-correlations of IGL and IGL with galaxy clustering, respectively. They are calculated using Eq.\,(\ref{eq:cxy}), and their power spectra are given by
\begin{equation}
\begin{aligned}
P_{{\rm \nu^h}{\rm g}}&=P^{\rm 1h}_{\rm \nu g}+P^{\rm 2h}_{\rm \nu g}\;,\\
P_{{\rm \nu^g}{\rm g}}&=P^{\rm 1h}_{\rm gg}+P^{\rm 2h}_{\rm gg}\;.
\end{aligned}
\end{equation}
The one-halo and two-halo terms of the IHL-galaxy power spectrum can be written as
\begin{equation}
\begin{aligned}
P^{\rm 1h}_{{\rm \nu^h}{\rm g}}=&\int \diff M\, \frac{\diff n}{\diff M}\, \frac{\left\langle N_{\rm{gal}}\right\rangle}{\bar{n}_{\rm{gal}}} f_{\rm IHL}(M)\,L(M,0)\,u^2(k | M)\;,\\
P^{\rm 2h}_{{\rm \nu^h}{\rm g}}=&P^{\rm lin}(k)\, \int \diff M\, \frac{\diff n}{\diff M}\, f_{\rm IHL}(M)\,L(M,0)\, b(M | z)\, u(k | M)\\
& \quad \times \int \diff M\, \frac{\diff n}{\diff M}\, \frac{\left\langle N_{\rm{gal}}\right\rangle}{\bar{n}_{\rm{gal}}}\, b(M | z)\,u(k | M)\;.
\end{aligned}
\end{equation}
We also present, in the lower panel of Fig.\,\ref{fig:cross_sp}, the angular power spectrum and correlation coefficient of the cross-correlation between EBL at 2.2\,$\si{\micron}$ and galaxy clustering in the tomographic redshift bin with $0.4\!<\!z_{\rm ph}\!<\!0.6$.

\section{Forecasts}
\label{analysis}
In this paper, we model all components of the EBL and use cross-correlations of multi-band intensity maps to separate the individual contributions. In particular, we focus on IHL and EoR components in order to obtain more precise constraints on the models, the fraction of total halo luminosity, and the star formation rate density. We generate mock SPHEREx and \Euclid data as follows. First, we generate fiducial auto- and cross-angular power spectra using the theoretical framework described in Sect.\,\ref{sc:model}. For the SPHEREx EBL, the fiducial power spectra are constructed based on a physically motivated multi-component model \citep{Cooray12,Helgason12,Zemcov14}, and combined with the instrumental characteristics of SPHEREx. For the Euclid cosmic shear and galaxy clustering, except for the cosmological parameters, all other inputs are taken from the FS2 results. Then, considering the instrumental and observational characteristics of SPHEREx and Euclid, we estimate the corresponding uncertainties on the power spectra through error analysis. We assume that these uncertainties are Gaussian and add random Gaussian errors to the fiducial spectra to generate the mock data. Then we analyse the data covariance matrix of the power spectrum, and used an MCMC method to constrain theoretical models. Through joint analysis of the SPHEREx EBL survey, \Euclid galaxy clustering, and cosmic shear surveys, we enhance constraints on various model parameters, facilitating the accurate separation of the IHL component from the EBL.

\subsection{SPHEREx survey}
\label{SPHEREx}
SPHEREx will be the first all-sky near-infrared spectral survey, creating a legacy archive of spectra with high sensitivity \citep{Dore14,Dore16,Dore18,Crill20,Bock26}. It includes six detector arrays that provide wide wavelength coverage and multi-band capabilities, across 102 spectral resolution elements from 0.75 to 5.0\,$\si{\micron}$ \citep{Crill20,Bock26}. This allows SPHEREx to obtain detailed spectral and image information in the near-infrared range. SPHEREx will observe two specific deep fields at the NEP and SEP. Each deep field covers approximately 100\,$\deg^2$ and reaches a limiting depth of AB magnitude 21.5--21.9 for $0.75\!<\!\lambda/\si{\micron}\!<\!3.8$ and 19.8--20.8 for $3.8\!<\!\lambda/\si{\micron}\!<\!5.0$ for point sources with $5\,\sigma$ detection \citep{Bock26}. These regions were selected to overlap with \Euclid, maximising the benefits of multi-band data for comprehensive scientific analysis. 

SPHEREx uses a dichroic beamsplitter to project the focal plane onto two $1\times3$ mosaics of H2RG HgCdTe charge-integrating detector arrays. These detectors capture sky images with an instantaneous field of view of $\ang{3.5;;} \times \ang{3.5;;}$ and a pixel resolution of $\ang{;;6.2}$, such that each $1\times3$ mosaic has a field of view of $\ang{3.5;;} \times \ang{11.3;;}$. Each detector has its own linear variable Ffilter (LVF), and each LVF contains 17 spectral resolution elements, making a total of 102 spectral resolution elements, with spectral resolution $\lambda/\Delta\lambda = 41$ in the range $0.75\!<\!\lambda/\si{\micron}\!<\!2.42$, $\lambda/\Delta\lambda = 35$ for $2.42\!<\!\lambda/\si{\micron}\!<\!3.82$, $\lambda/\Delta\lambda = 110$ for $3.82\!<\!\lambda/\si{\micron}\!<\!4.42$, and $\lambda/\Delta\lambda = 130$ for $4.42\!<\!\lambda/\si{\micron}\!<\!5.0$ \citep{Bock26}. We note that there is an overlap of a quarter of a spectral resolution element between adjacent bands and the dichroic. To enhance the sensitivity of SPHEREx, we divided the entire wavelength range into ten bands, ensuring that the spectral resolution satisfies $4.5 \leq \lambda / \Delta\lambda \leq 8.1$. For the four detectors with shorter wavelength ranges, each detector's wavelength range is split into two bands, while the coverage of the two longer wavelength detectors occupies one band each. The parameters of the ten SPHEREx bands are listed in Table\,\ref{tab:bands}.
\newcommand{\pd}{\phantom{1}}
\newcommand{\pp}{\phantom{-}}
\begin{table}
\centering
\caption{\label{tab:bands} Parameters of SPHEREx bands.}
\begin{tabular}{ccccc}
\hline\hline
\rule[-2mm]{0mm}{6mm}
Band & $\lambda_{\rm c}$ & $\Delta\lambda$ & $\rm \lambda/\Delta\lambda$ & $N_{\ell}$ \\
& ($\si{\micron}$) & ($\si{\micron}$) & &($\rm nW^2\,m^{-4}\,sr^{-2}$)\\
\hline
\rule[-2mm]{0mm}{6mm}
1 & 0.82 & 0.16 & 5.2 & $1.46\times10^{-9\pd}$\\
2 & 0.99 & 0.19 & 5.2 & $1.09\times10^{-9\pd}$\\
3 & 1.24 & 0.24 & 5.2& $6.70\times10^{-10}$\\
4 & 1.50 & 0.29 & 5.2& $4.29\times10^{-10}$\\
5 & 1.83 & 0.35 & 5.2& $2.80\times10^{-10}$\\
6 & 2.22 & 0.43 & 5.2& $1.65\times10^{-10}$\\
7 & 2.75 & 0.62 & 4.5& $7.60\times10^{-11}$\\
8 & 3.44 & 0.78 & 4.5& $4.53\times10^{-11}$\\
9 & 4.13 & 0.60 & 6.9& $1.02\times10^{-10}$\\
10 & 4.72 & 0.58 & 8.1& $2.26\times10^{-10}$\\
\hline
\end{tabular}
\end{table}

We assume that the noise fluctuation in the pixels satisfies a typically Gaussian distribution, and the noise characteristics of the images in each wavelength element are similar. Thus we can combine the images by taking a pixel-wise average model. The noise variance in the combined image is the average of the noise variances from each wavelength element, $\delta^2_{\rm bnad} =N^{-2} \sum^{N}_{i=1}\delta^2_{\lambda_i}$, where $\delta_{\lambda}$ is the sensitivity per pixel in the SPHEREx Deep surveys. Therefore, the instrument of noise can be estimated by $\Omega_{\rm pixel}\delta^2_{\rm band}$, where $\Omega_{\rm pixel}$ is the pixel area. The instrument noise of the ten SPHEREx bands are listed in Table\,\ref{tab:bands}. We can mask the bright galaxies to remove their influence on the EBL. Galaxies fainter than AB magnitude 20.5 can have a significant impact on EBL, particularly at infrared and optical wavelengths. We estimate the contribution of unresolved sources to the EBL by setting the upper limit of the IGL integration to AB\,=\,20.5 at 2.2\,$\si{\micron}$. Zodiacal light fluctuations are assumed smooth and sub-dominant in the NEP and SEP deep fields.

The statistical error $\delta C(\ell)$ of the auto-angular power spectrum can be written as \citep{Hu02,Zemcov14,Mitchell15}
\begin{equation}
\delta C(\ell)=\sqrt{\frac{2}{f_{\rm sky}\,(2 \ell+1)\,\Delta \ell}}\,\left[C(\ell)+N(\ell)\right]\;,
\label{eq:delta_C}
\end{equation}
where $f_{\rm sky}$ is the fraction of the total sky covered by the deep field, including the contribution of all masks, $\Delta \ell$ is the width of the $\ell$ bin, $C(\ell)$ is the total angular power spectrum of all components, and $N(\ell)$ is noise. For the cross-correlation, the statistical error $\delta C^{\,ij}(\ell)$ becomes
\begin{equation}
\delta C^{\,ij}(\ell)=\sqrt{\frac{\left[C^i(\ell)+N^i(\ell)\right]\,\left[C^j(\ell)+N^j(\ell)\right]+\left[C^{\,ij}(\ell)\right]^2}{f_{\rm{sky}}\,(2\ell+1)\, \Delta\ell}}\;,
\end{equation}
where $C^i(\ell)$ and $C^j(\ell)$ are the auto-angular power spectra, $N^i(\ell)$ and $N^j(\ell)$ are the noise from $i$th and $i$th bands, respectively, and $C^{\,ij}(\ell)$ is the cross angular power spectrum.
 
\begin{figure*}[htbp!]
\centering
\includegraphics[width=1.0\textwidth]{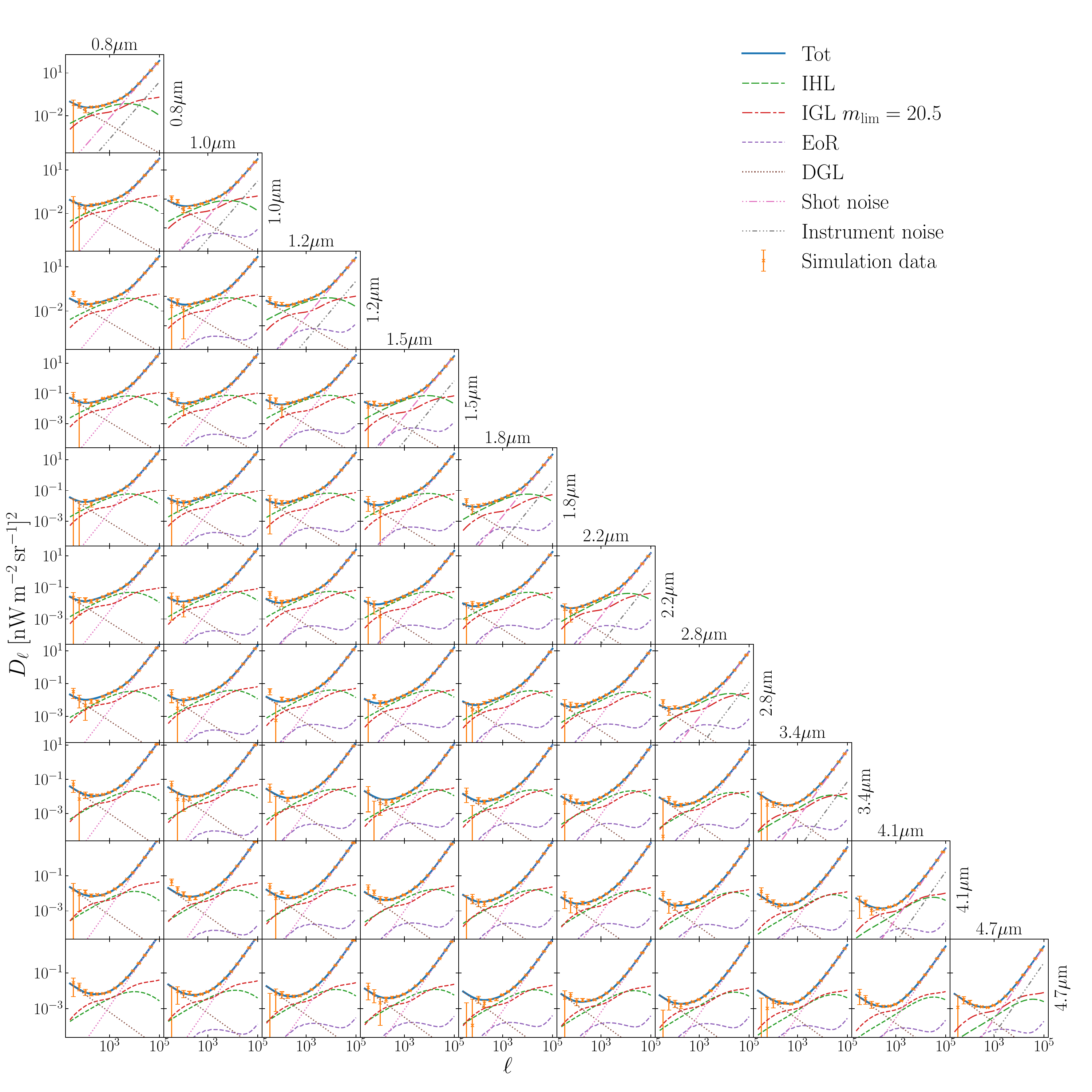}
\caption{Simulation data and theoretical models of the auto- and cross-power spectra of all ten bands. The orange data points with error bars are the simulation data. The blue solid line denotes the total angular power spectrum with noise; the green, red, and purple dashed lines denote the power spectra of IHL, IGL, and EoR signal, respectively. DGL is shown with a brown dotted line; shot noise and instrument noise are shown with pink and grey dash-dotted lines, respectively. The orange crosses with error bars represent mock data.}
\label{fig:EBL_all}
\end{figure*}

The minimum multipole $\ell_{\rm min}$ of the angular power spectrum is related to the observed sky area. For the SPHEREx deep field, the minimum multipole of the angular power spectrum is approximately $\ell_{\rm min} = 32$. the maximum multipole $\ell_{\rm max} = 100\,000$ is related to the pixel scale, also including the effective beam window for each band. We divide our data into 15 log-spaced multipole bins from $\ell_{\rm min}$ to $\ell_{\rm max}$, and create the simulation data by adding Gaussian errors and present them using orange data points with error bars in Fig.\,\ref{fig:EBL_all}. 

\subsection{\Euclid survey}
\label{Euclid}
\Euclid is a medium-class space mission of the European Space Agency (ESA), which will conduct photometric and spectroscopic galaxy surveys over approximately 14\,000\,$\deg^2$ of the extragalactic sky \citep{EuclidSkyOverview}. The photometric survey will measure the positions and shapes of over a billion galaxies, enabling analyses of weak lensing and galaxy clustering. In this work, we focus on the EDFN and EDFS fields.
To study the cosmic shear and galaxy clustering for \Euclid, we first need to find a galaxy catalogue that can represent the surveys. Therefore, we create a \Euclid sample using the \Euclid flagship simulations~2 (FS2, \citealt{EuclidSkyFlagship}), which was retrieved from CosmoHub \citep{Carretero17,Tallada20}.

FS2 is a high-resolution $N$-body simulation designed to support the scientific goals of the \Euclid survey. It covers a cubic volume with a side length of 3600\,\si{\hMpc} and includes over 4 trillion particles, each with a mass of $10^9\,h^{-1}\,\si{\solarmass} $. This simulation accurately resolves dark matter haloes with masses as low as $10^{11}\, h^{-1}\,\si{\solarmass}$, which host the faintest galaxies observable by \Euclid. The simulation is based on a flat $\Lambda$CDM cosmology with matter density $\Omega_{\rm m} = 0.319$, baryon density $\Omega_{\rm b} = 0.049$, and dark energy density $\Omega_{\Lambda} = 0.681 - \Omega_{\rm r}-\Omega_{\rm \nu}$, with a photon density parameter $\Omega_{\rm r}$ = 0.00005509, and a neutrino density parameter $\Omega_{\rm \nu}$ = 0.00140343 corresponding to the minimal neutrino mass. Other key parameters include a Hubble constant $h=0.67$, scalar spectral index $n_{\rm s}=0.96$, and primordial power spectrum amplitude $A_{\rm s}=2.1\times 10^{-9}$. The FS2 contains 3.4 billion galaxies in a magnitude-limited sample with $\HE<26$.

\begin{figure}[htbp]
\centering
\includegraphics[width=1.0\columnwidth]{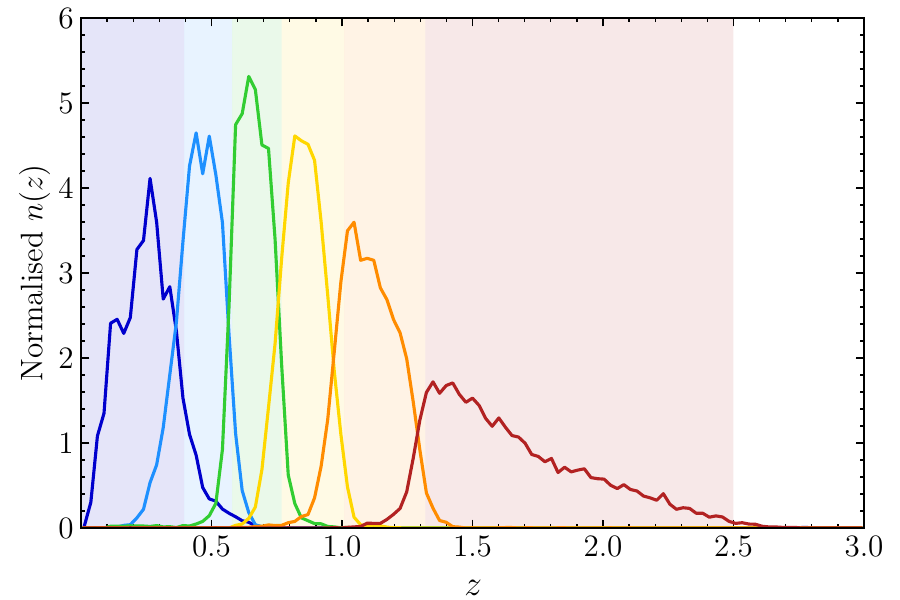}
\caption{Redshift distribution of \Euclid galaxies in the FS2. The shaded regions indicate the six tomographic bins with equal galaxy number densities $\bar{n}=4.29$ galaxies per arcmin$^2$. The real redshift distributions of the six bins are shown by the coloured lines.}
\label{fig:redshift} 
\end{figure}

The photometric redshifts of galaxies were obtained using the nearest-neighbour photometric redshift pipeline, which also estimates the redshift probability distributions of galaxies \citep{Desprez-EP10}. To extract more information from the cosmic shear and galaxy clustering, we divide the redshift range into six different photometric redshift bins and study the angular power spectra of these tomographic bins. We select galaxies with a photometric redshift of less than 2.5, and show the redshift distributions of the six bins in Fig.\,\ref{fig:redshift}. The coloured shaded regions indicate the six bins with equal galaxy number densities $\bar{n}=4.29$ galaxies per arcmin$^2$, leading to a total of roughly 25.7 galaxies per arcmin$^2$. The solid coloured lines are the true redshift distribution $n(z)$ of the galaxies whose photometric redshift are measured within the tomographic redshift bins. We show the kernels of gravitational shear and galaxy on the right side of Fig.\,\ref{fig:kernel}. We find that the overlap fractions between adjacent bins are about 30\%, 14\%, 19\%, 16\%, and 9\%, respectively, while the overlap fractions between non-adjacent bins are all below 1\%. This leads to weak cross-correlations between these bins and implies an increase in the shot-noise level by a factor of more than 5 for all cross-bin spectra except that between the first two bins. Therefore, we only neglect the cross-power spectra of galaxy clustering between different bins. 

To avoid deviation from the Limber approximation \citep{Limber53,LoVerde08}, we considered the minimum multipole to be $\ell_{\rm min} = 70$. For cosmic shear, we consider two scenarios for the maximum multipole cut $\ell_{\rm max}$ given by \cite{Blanchard-EP7}, with optimistic and pessimistic values for $\ell_{\rm max}$ of 3000 and 750, respectively. Similarly, we take $\ell_{\rm max}$ of 5000 and 1500 for galaxy clustering as optimistic and pessimistic values, respectively. We bin our data in multipoles with 12 bins for cosmic shear and galaxy clustering in log-spaced multipole bins in the range from $\ell_{\rm min}$ to $\ell_{\rm max}$. We take the overlapping edges of the multipole range of the auto-power spectra as the maximum and minimum multipole for the cross-correlation.

\subsection{MCMC fitting}
\label{fitting}
After obtaining the angular power spectra, we perform joint fits for the SPHEREx EBL, \Euclid cosmic shear, and galaxy clustering using the MCMC method \citep{Metropolis53}. We use \texttt{emcee} \citep{Foreman13} to sample the posterior and analyse constraints on model parameters. In this work, we estimate the likelihood function by analysing the $\chi^2$ distribution $\mathcal{L} \propto \exp \left(-\chi_{\rm tot}^2 / 2\right)$, where the total $\chi_{\rm tot}^2$ for joint surveys is given by $\chi_{\rm tot}^2=\chi_{\nu}^2+\chi_{\gamma}^2+\chi_{\rm g}^2+\chi_{\nu \gamma}^2+\chi_{\rm \nu g}^2$, where $\chi_{\nu}^2$, $\chi_{\gamma}^2$, $\chi_{\rm g}^2$, $\chi_{\nu \gamma}^2$, and $\chi_{\rm \nu g}^2$ are for the EBL, cosmic shear, galaxy clustering, EBL-cosmic shear, and EBL-galaxy clustering power spectra, respectively. The $\chi^2$ can be calculated as
\begin{equation}
\chi^2=\sum_{ij}^N\left(C_{\rm{obs}}^i-C_{\rm{th}}^i\right) \,\operatorname{Cov}_{ij}^{-1}\,\left(C_{\rm{obs }}^j-C_{\rm{th}}^j\right)\;,
\end{equation}
where $C_{\rm obs}$ and $C_{\rm th}$ are the observed and theoretical power spectra, respectively, and $\operatorname{Cov}_{ij}^{-1}$ is the inverse of the covariance matrix.

We assume that the model parameters of IHL, $\alpha$, depend on redshift, while $A_{\rm IHL}$ and $\beta$ depend on halo mass. We divide the range of redshift and halo mass into four and seven tomographic bins, respectively. The redshift bins have equal widths, while the last six halo mass bins are uniformly distributed on a logarithmic scale and the first halo mass bin twice as wide as the others. Therefore, we need to fit four $\alpha$, seven $\beta$, and seven $A_{\rm IHL}$ values for the IHL model. 

For IGL and shot noise, we fix the evolution parameters of the luminosity function to the best-fit values obtained by \cite{Helgason12} and add an amplitude $A^{\phi}$ as a free parameter for each band. For EoR, we divide the $f_*(z)$ into five tomographic redshift bins, with the boundaries of these bins being $\{ 6.0,\,7.14,\, 9.17,\, 11.34,\, 14.03,\, 30.0\}$. Since the signal from the highest-redshift bin is much weaker than that from the others, we only fit the star formation efficiency $f_*$ for the four lower-redshift bins. For simplicity, the fiducial amplitude $A_{\rm DGL}$ follows \cite{Sano17}. $A_{\rm DGL}$ can be considered as a wavelength-dependent function. We add an amplitude $A^{\prime}_{\rm DGL}$ for this function as a free fitting parameter. 

Altogether, the theoretical model contains 50 free parameters, including 18 IHL parameters: four $\alpha$, seven $\beta$, and seven $A_{\rm IHL}$; ten IGL amplitudes $A^{\phi}$; ten shot-noise amplitudes $A_{\rm SN}$; four star formation efficiencies $f_*$; two IA model parameter $A_{\rm IA_1}$ and $A_{\rm IA_2}$; and six multiplicative calibration parameters $m_i$. In this work, we set flat priors: $\alpha \in[-3,3]$; $\beta \in [-1,10]$; $\logten(A_{\rm IHL}) \in [-10,0]$; $A^{\phi} \in [0,2]$; $A_{\rm SN} \in [0,2]$; $f_*\in [0,0.1]$; $A_{\rm DGL} \in [0,2]$ and $A_{\rm IA} \in [0,2]$. 

\begin{figure}[htbp]
\centering
\includegraphics[width=1.0\columnwidth]{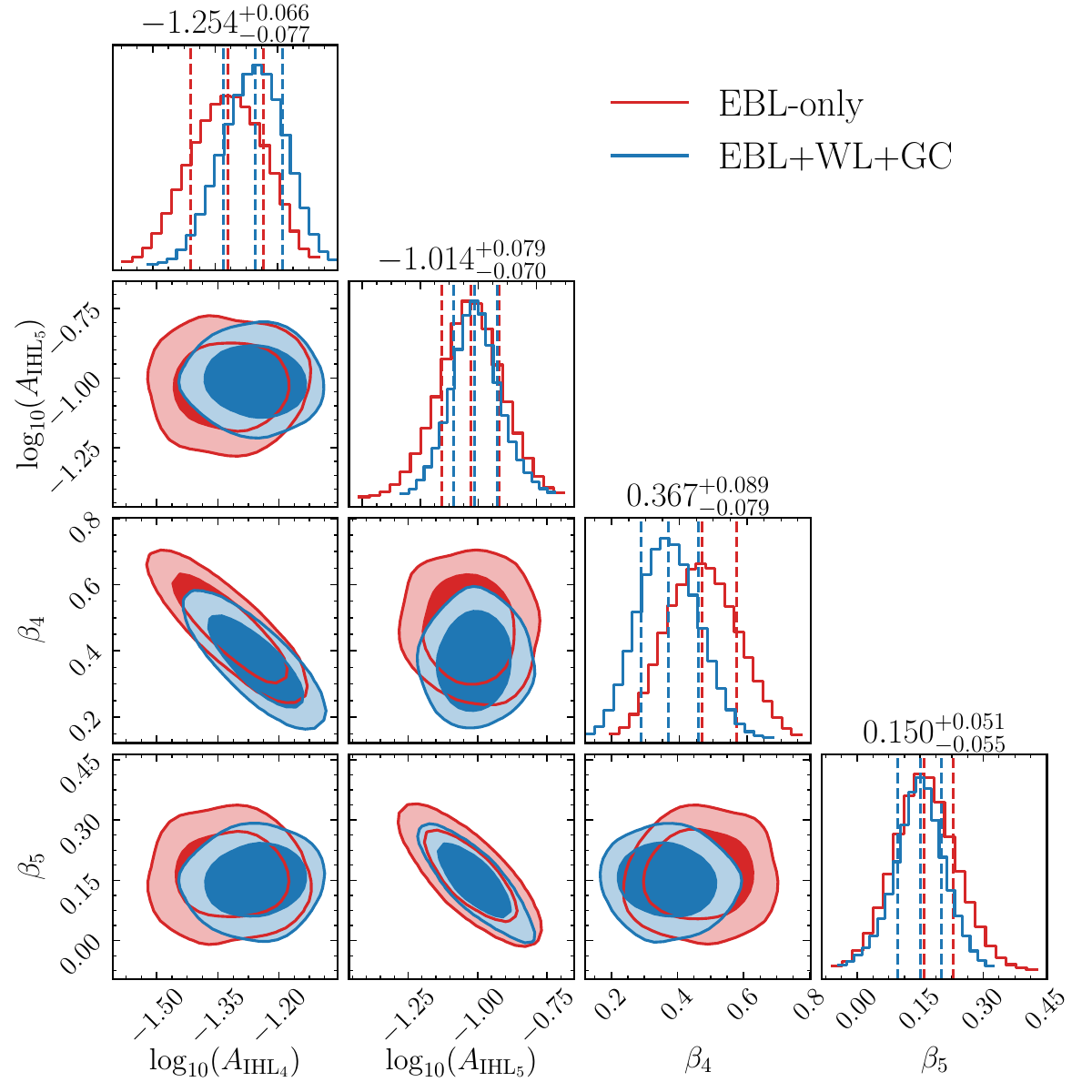}
\caption{Contour maps with $1\,\sigma$ (68.3\%) and $2\,\sigma$ (95.5\%) confidence levels of the $\beta$ and $A_{\rm IHL}$. The red contours denote the results of the SPHEREx EBL-only, and the blue contours are the joint constraint results.}
\label{fig:corner} 
\end{figure}

For the multiplicative calibration parameters $m_i$, we adopt a Gaussian prior $\mathcal{N}\left(0, 0.003^2\right)$, where $\sigma_{\rm m} \le 0.003$ can be achieved in stage IV surveys such as the \Euclid and LSST surveys. We consider four cases: (i) EBL-only, which includes the SPHEREx EBL auto- and cross-spectra; (ii) EBL+WL, which includes the SPHEREx EBL auto- and cross-spectra, the \Euclid shear auto- and cross-spectra, and the cross-spectra between SPHEREx EBL and \Euclid shear; (iii) EBL+GC, which includes the SPHEREx EBL auto- and cross-spectra, the \Euclid galaxy auto-spectra, and the cross-spectra between SPHEREx EBL and \Euclid galaxy clustering; (iv) EBL+WL+GC, which includes all of the above components. The parameters of these surveys are listed in Table\,\ref{tab:cases}. We used 15 log-spaced multipole bins for the EBL spectra and 12 for the EBL–cosmic shear, cosmic shear, EBL–galaxy clustering, and galaxy clustering spectra. Since $A_{\rm IA}$ and $m_i$ appear only in the cosmic shear-related spectra, combining all auto- and cross-spectra, we obtain the number of degrees of freedom as $N_{\rm dof}=783$ for EBL-only, $N_{\rm dof}=1747$ for EBL+WL, $N_{\rm dof}=1575$ for EBL+GC, and $N_{\rm dof}=2539$ for EBL+WL+GC. We use the MCMC method to constrain these free parameters in the theoretical model and calculate the acceptance probability of the chain points. For each case, we run 160 walkers. After reaching convergence \citep{Gelman92}, each chain obtained 50\,000 sample points. We then perform burn-in and thinning processes on the data and combine all chains, resulting in a total of 10\,000 chain points. We show the fitting results and 68\% confidence levels in Table\,\ref{tab:result}. In the MCMC analysis, we find that there is a noticeable correlation of the same IHL model parameters across adjacent bins. In the case of the IHL model, the $\beta$ and $A_{\rm IHL}$ within the same halo mass bin are strongly negatively correlated, as a mass power-law index can be partially compensated by a lower amplitude value. In Fig.\,\ref{fig:corner}, we show the two-dimensional contours for $A_{\rm IHL}$ and $\beta$ in two halo mass bins. In each halo-mass bin, a clear negative correlation between $A_{\rm IHL}$ and $\beta$ is observed.
\begin{table*}
\centering
\caption{\label{tab:cases} Survey parameters.}
\begin{tabular}{ccccccc}
\hline\hline
\rule[-2mm]{0mm}{6mm}
Component & \multicolumn{2}{c}{Coverage area} & $f_{\rm sky}$ & $\ell_{\rm min}$ & $\ell_{\rm max}$ & Number of bins \\
& North & South & &\\
\hline
\rule[-2mm]{0mm}{6mm}
EBL auto/cross-correlation & $100\,\deg^2$ & $100\,\deg^2$ & 0.00485 & 32 & 100\,000 & 15\\
Shear auto/cross-correlation & $20\,\deg^2$ & $23\,\deg^2$ & 0.00104 & 70 & 3000 & 12\\
Galaxy auto-correlation & $20\,\deg^2$ & $23\,\deg^2$ & 0.00104 & 70 & 5000 & 12\\
EBL $\times$ shear & $20\,\deg^2$ & $23\,\deg^2$ & 0.00104 & 70 & 3000 & 12\\
EBL $\times$ galaxy & $20\,\deg^2$ & $23\,\deg^2$ & 0.00104 & 70 & 5000 & 12\\
\hline
\end{tabular}
\end{table*}

\begin{figure*}[htbp!]
    \centering
    \includegraphics[width=0.49\textwidth]{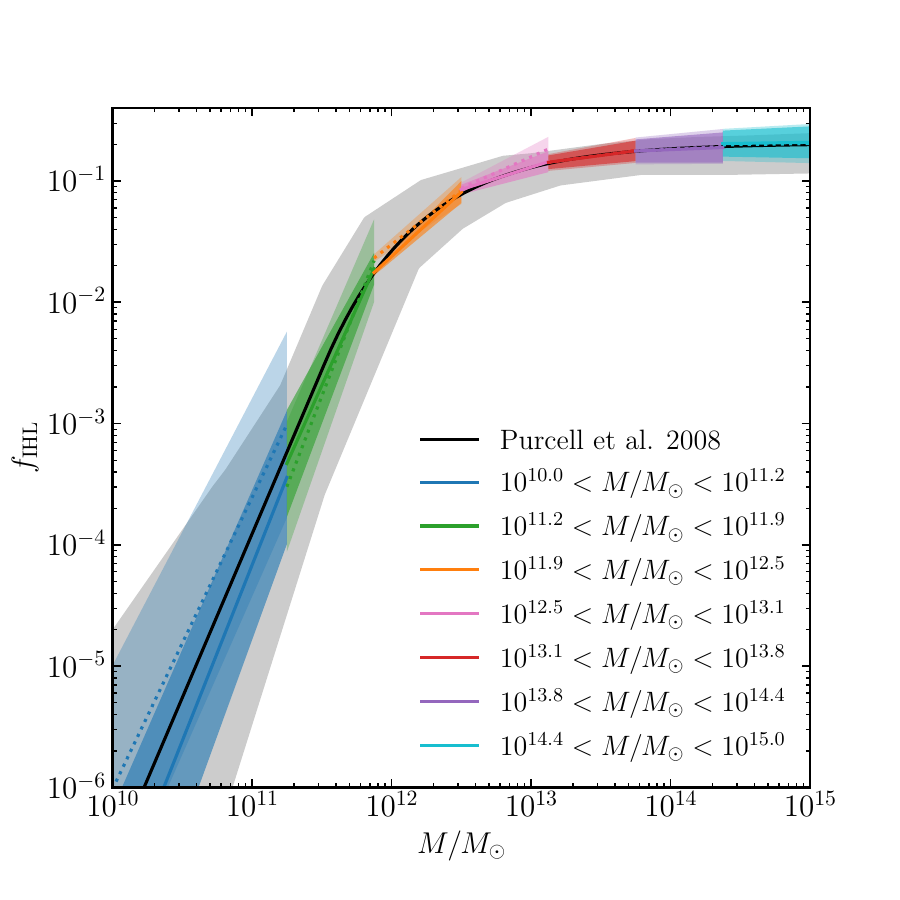}
    \includegraphics[width=0.49\textwidth]{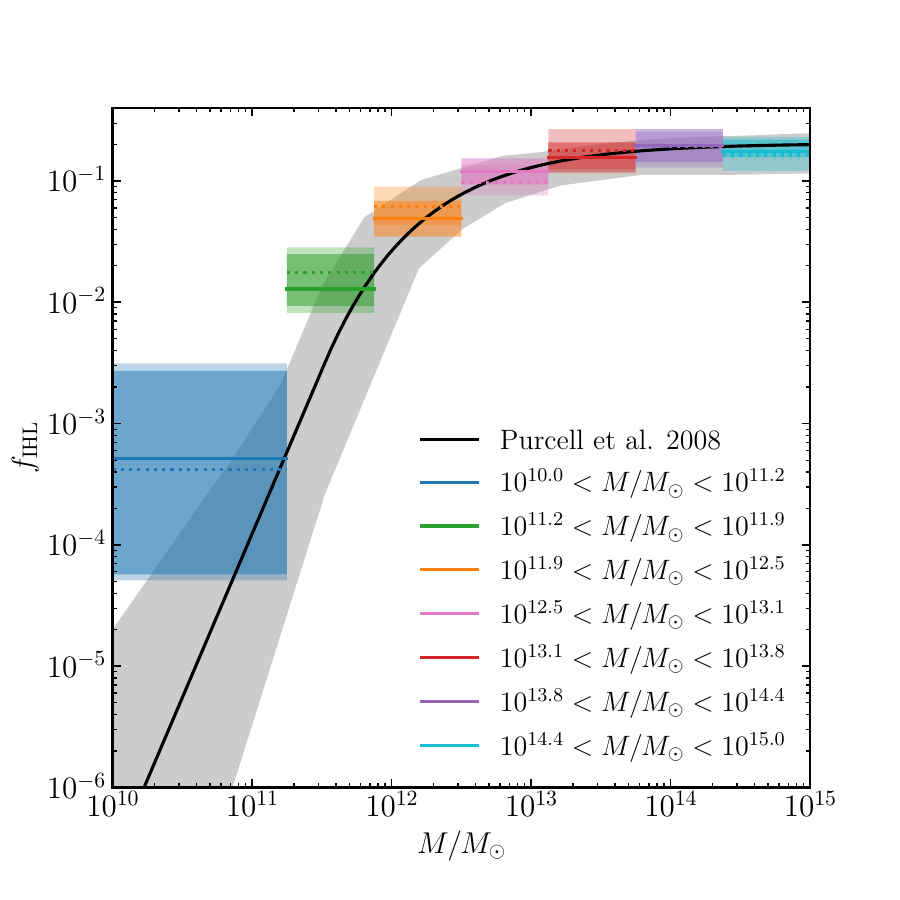}
    \caption{Fraction of total halo luminosity. The halo mass is divided into seven bins ranging from $10^{10}$ to $10^{15}\,\si{\solarmass} $, the last six bins are uniformly distributed on a logarithmic scale and the first bin is twice as wide as the others. \emph{Left}: Results using a power-law fitting. \emph{Right}: Results using a constant value for $f_{\rm IHL}$. The black line represents the fiducial IHL fraction, which is based on the study by \cite{Purcell08}, and the grey shaded region contains 68 per cent of the distribution. The coloured lines and shaded regions are the best-fit results from the MCMC fitting and the $1\,\sigma$ confidence level constraints, respectively. The coloured dotted lines with light shaded regions and coloured solid lines with dark shaded regions are the fitting results using SPHEREx EBL-only (EBL-only) and SPHEREx EBL + \Euclid cosmic shear + \Euclid galaxy clustering (EBL+WL+GC), respectively.}
    \label{fig:f_ihl}
\end{figure*}

\begin{figure}[htbp]
\centering
\includegraphics[width=1.\columnwidth]{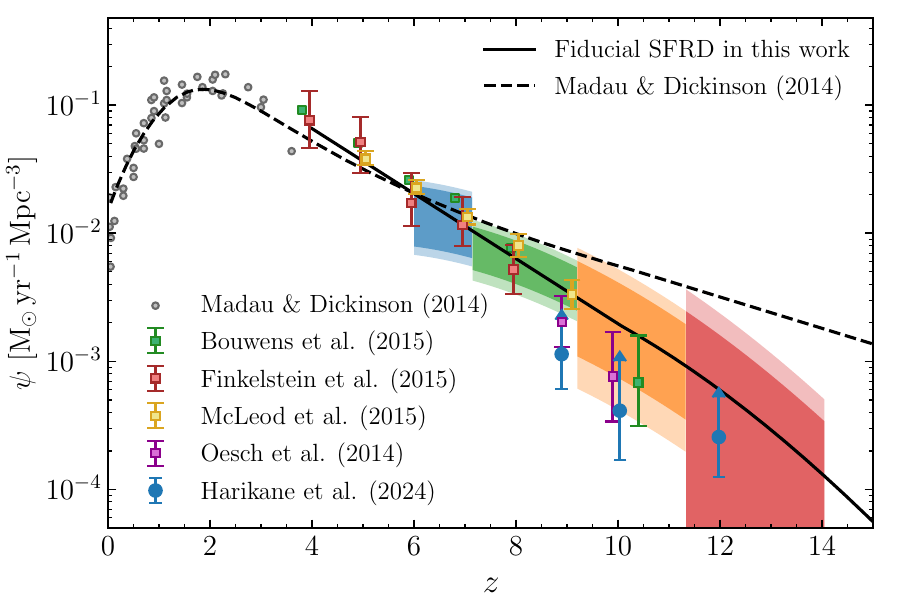}
\caption{Star formation rate density (SFRD). The solid line is the fiducial SFRD, which is fit to the total SFRD from \cite{Finkelstein16} at $z \leq 10$, and assumes $f_*=0.01$ at $z>10$. The dashed line is from \cite{Madau14}. The light and dark shaded regions show respectively the results of fitting with EBL-only and EBL+WL+GC, as discussed here.}
\label{fig:SFRD}
\end{figure}

We find that all the fiducial values fall within the $1\,\sigma$ confidence level of posteriors. For the joint survey combining SPHEREx EBL, \Euclid cosmic shear, and \Euclid galaxy clustering, the constraints on the model parameters are significantly improved compared to EBL-only, leading to enhanced accuracy in the parameter estimation. Relative to the SPHEREx EBL-only case, the EBL+WL and EBL+WL+GC analyses reduce the $1\,\sigma$ uncertainties of the IHL parameters by 10--20\% and 10--30\%, respectively. For the EoR model, the constraints on $f_*$ are improved with uncertainty reductions of about 10--20\% for EBL+WL and up to 20--30\% for EBL+WL+GC. Using the best-fit model and uncertainties as determined by MCMC model fits, we convert the parameter constraint to a measure of the IHL luminosity and SFRD. For the IHL fraction $f_{\rm IHL}$, we also considered a simple fitting model, where the value of $f_{\rm IHL}$ is treated as a constant in each halo mass bin. In Fig.\,\ref{fig:f_ihl}, we show the fiducial IHL fraction and MCMC constraint results. The left panel shows the results using a power-law fit, and the right panel shows the results when assuming a constant value for $f_{\rm IHL}$. We divide the IHL fraction $f_{\rm IHL}$ into seven halo mass bins from $10^{10}$ to $10^{15}\,\si{\solarmass}$, where the last six bins are uniformly distributed on a logarithmic scale, and the first bin is twice as wide as the others. The black line represents the fiducial IHL fraction, which is based on the study by \cite{Purcell08}, and the grey shaded region contains 68 per cent of the distribution. The coloured lines and shaded regions are the best-fit results from the MCMC fitting and the $1\,\sigma$ confidence level constraints, respectively. The coloured dotted lines with light shaded regions and coloured solid lines with dark shaded regions are the fitting results using EBL-only and EBL+WL+GC, respectively. Compared to the referenced IHL fraction, our results indicate a significant improvement, especially in the halo mass range of $1.58 \times 10^{11}< M/\si{\solarmass}<6.31 \times 10^{13}$, where the $1\,\sigma$ confidence level enhancement exceeds 50\%. The joint constraints reduce the error bar by approximately 5--30\% compared to only using SPHEREx EBL data.
Since the EoR signal comes from high-redshift $z\!\ge\!6$, and the redshifts of galaxy samples observed by \Euclid are significantly lower than the EoR, the cross-correlation between them is expected to be negligible. There can be residual lensing of high-$z$ sources by low-$z$ structure, but the source kernel and tomographic bins used here suppress it, giving negligible correlation between the EoR signal and \Euclid cosmic shear (or galaxy clustering). Therefore, the cross-correlation between \Euclid and SPHEREx is more beneficial for separating the EoR signal from the EBL. The predicted results for the SFRD measurements are presented in Fig.\,\ref{fig:SFRD}. The solid line is the input reference SFRD, which is fitted to the total SFRD form \cite{Finkelstein16} at $6 < z < 10$, and set to $f_* = 0.01$ at $z \ge 10$. The light and dark shaded regions in our redshift bins indicate the $1\,\sigma$ error bounds for EBL-only and EBL+WL+GC, respectively. We find that the fiducial SFRD lies within the $1\,\sigma$ uncertainty on the SFRD measurements, and our method provides strong constraints, with the highest precision achieved at $7.14\!<\!z\!<\!9.17$. Compared to the EBL-only case, the joint constraints improve the measurement precision by 15--30\%, with the largest gain obtained at $9.17\!<\!z\!<\!11.34$. The grey circles at $z\!<\!4$ are from the compilation by \cite{Madau14}, while the green, red, yellow, purple squares and blue dots are constraints from \cite{Bouwens15}, \cite{Finkelstein15}, \cite{ McLeod15}, \cite{Oesch14}, and \cite{Harikane24} respectively. The measurements in \cite{Harikane24} are based on an analysis of JWST data. Compared to these studies, our approach allows for SFRD measurements at higher redshifts, providing stronger constraints at $6\!<\!z\!<\!11.3$. The forecasted constraints on the EoR related parameters depend on the assumed fiducial SFRD model. A higher fiducial SFRD leads to a larger EoR contribution to the EBL fluctuations and hence a higher effective signal-to-noise ratio for the EoR component. The resulting parameter uncertainties therefore depend on the assumed fiducial model.

\section{Discussion}
\label{discussion}
We performed a series of internal checks to assess the stability of our results:
(i) doubling the DGL amplitude within the prior range shifts the large-scale EBL auto-power but has negligible impact on the cross-spectra-driven constraints; (ii) modest variations in the masking threshold (equivalently, the shot-noise amplitude prior) primarily exchange power between IGL and shot noise, while leaving IHL parameters stable once cross-spectra are included; and (iii) recovering all injected parameters within $1\,\sigma$ across four data-vector choices (EBL-only; EBL+WL; EBL+GC; EBL+WL+GC) provides a strong end-to-end validation.

\subsection{Limitations}
Our current forecasts adopt several simplifying assumptions that can be relaxed in future work. We model IA with two TATT amplitude and do not include an explicit redshift or luminosity dependence; we also do not propagate photometric redshift bias and scatter in the shear and clustering kernels beyond what is implicit in the FS2-like mocks. The IGL term relies on a Schechter-function evolution and a simple HOD; departures in the faint-end slope or satellite occupation at the masking limit could shift the balance between IGL and shot noise. Finally, our DGL model assumes a single power-law slope at high latitude; spatial variations in dust decorrelation across wavelengths could introduce additional large-scale covariance that we will include in future studies.

\subsection{Outlook}
The analysis strategy demonstrated here -- spectral EBL mapping cross-correlated with tomographic cosmic shear and galaxy density -- will scale naturally with forthcoming data. On the SPHEREx survey side, improved deep-field maps and refined masking will lower the effective shot-noise floor and sharpen the IGL--IHL separation. On the \Euclid survey side, shear calibration, intrinsic-alignment modelling, and photo-$z$ characterisation from the full survey will tighten priors and reduce degeneracies. Beyond \Euclid, cross-correlations with Rubin/LSST and CSST will further enhance tomographic constraints. Incorporating more flexible IA models, explicit photo-$z$ nuisance parameters, and a wavelength-dependent DGL template into the joint fit are natural next steps. With these extensions, SPHEREx$\times$\Euclid is poised to deliver component-resolved near-IR EBL measurements that inform the build-up of intra-halo stars, the abundance of faint galaxies below current detection thresholds, and the star formation history deep into the EoR.

\section{Summary and conclusion}
\label{conclusions}
We have developed and tested a joint analysis framework that combines multi-band SPHEREx intensity mapping with \Euclid cosmic shear and galaxy-clustering tomography to separate the principal contributors to EBL fluctuations. Our forward model includes IHL, unresolved IGL below a masking threshold (AB\,=\,20.5 in the deep fields), and a high-$z$ EoR component based on Pop~II/III emissivities, together with DGL and shot noise. On the lensing side we implement tomographic shear spectra with multiplicative calibration parameters per bin and TATT IA. Using mock data that match SPHEREx’s ten band coverage over the NEP deep fields ($\sim$\,100\,deg$^{2}$ total) and \Euclid’s six tomographic bins to $z=2.5$, we fit the combined data vector (EBL auto-spectra, shear and clustering auto-spectra, and all EBL$\times$shear and EBL$\times$galaxy cross-spectra) with an MCMC method.

Our main findings are as follows:
\begin{enumerate}
\item {Component-resolved EBL constraints.}
The joint analysis recovers all injected (fiducial) parameters within $1\,\sigma$, and \emph{reduces marginalised uncertainties on IHL parameters by 10--30\%} relative to EBL-only fits. The EoR star formation efficiency parameters improve by \emph{20--30\%}. These gains come primarily from the additional large-scale-structure information in the cross-spectra, which break degeneracies among IHL, IGL, and shot noise that are otherwise difficult to disentangle with intensity auto-spectra alone.
\item {IHL-halo connection.}
By allowing both a redshift-dependent IHL evolution index $\alpha(z)$ and a mass-dependent fraction $f_{\rm IHL}(M)$, we measure an IHL fraction that is significantly tightened across $10^{11.2}$--$10^{13.8}\,\si{\solarmass}$. Relative to SPHEREx-only, error bars on $f_{\rm IHL}$ shrink by 5--30\% in these mass ranges, and the best-fit trend remains consistent with a smoothly varying, approximately power-law dependence on halo mass.
\item {Cross-correlations as a lever arm.}
The pattern of EBL$\times$shear correlations is a strong discriminator: Within the adopted model framework, IHL exhibits a higher correlation coefficient with shear than IGL at fixed wavelength and multipole, consistent with diffuse intrahalo starlight tracing the gravitational potential more directly than sub-threshold galaxies selected by luminosity. Conversely, the EoR contribution shows negligible correlation with \Euclid shear and clustering given the limited kernel overlap, which helps isolate its (high-$z$) signal in the EBL bands.
\item {SFRD at high redshift.}
Translating the EoR constraints into star formation rate density, the joint analysis extends competitive SFRD constraints to $z\!\approx\!11$, with improved precision in the range $6\!\lesssim\!z\!\lesssim\!11.3$. While the absolute normalisation depends on the assumed Pop~II/III templates and escape fractions, the multi-band spectral shape and the weak cross-correlation with shear together stabilise the inference against low-$z$ contaminants.
\item {Control of systematic effects via cross-spectra.}
Foregrounds and instrument terms that dominate single-probe auto-spectra are suppressed or nulled in cross-correlation. In particular, DGL power that scales as a steep multipole law at high Galactic latitude, residual zodiacal fluctuations, and instrument noise carry negligible correlation with shear and thus do not bias the EBL$\times$shear mean. On the lensing side, we marginalise over per-bin multiplicative calibration parameters $m_i$ and amplitude $A_{\rm IA_1}$ and $A_{\rm IA_2}$ without erasing the astrophysical gains from cross-correlations.
\end{enumerate}

In summary, multi-band SPHEREx intensity mapping combined with \Euclid cosmic shear and clustering tomography provides a robust, systematics-aware pathway to decomposing the near-IR EBL. In our end-to-end forecasts, cross-correlations reduce marginalised errors on IHL by 10--30\% and on EoR star formation efficiency by 20--30\% relative to EBL-only, tighten the inferred $f_{\rm IHL}(M)$ across group-scale haloes, and extend SFRD constraints to $z\!\approx\!11$. These results make a strong case for treating EBL$\times$shear and EBL$\times$galaxy cross-spectra as core observables in the SPHEREx deep fields and a key probing for learning about diffuse stellar components and early galaxy formation.

\begin{acknowledgements}
Y.C. and A.C. thank the NASA SPHEREx science programmes for support. T.L. acknowledges funding from a GAANN Fellowship from the US Department of Education. We also acknowledge support from NASA grants NNX16AJ69G and NNX16AF39G. 
We thank Jordan Mirocha for helpful discussions and comments.
\AckEC
\AckCosmoHub
\end{acknowledgements}

\bibliographystyle{aa}
\bibliography{Euclid,SPHERExxEuclid}

\begin{appendix}
  \onecolumn 
  
\section{The best fits of the free parameters from the MCMC fitting.}
This appendix lists the best-fit values and corresponding uncertainties of the free parameters obtained from the MCMC analysis of the multi-component model.

\renewcommand{\arraystretch}{1.55}
\begin{longtable}{ccccccc}
\caption{\label{tab:result} Best fits and errors of the free parameters in the model from the MCMC fitting.}\\
\hline
\vspace{-5.5mm}\\
\hline
Parameter & Fiducial value & EBL-only & EBL+CS & EBL+GC & EBL+CS+GC & Prior (min,max)\\
\hline
\endfirsthead
\caption{continued.}\\
\hline
\vspace{-5.5mm}\\
\hline
Parameter & Fiducial value & EBL-only & EBL+CS & EBL+GC & EBL+CS+GC & Prior (min,max)\\
\hline
\endhead
\multicolumn{7}{c}{IHL}\\
$\alpha_1$ &$-1.05$ & $-1.058_{-0.033}^{+0.034}$ & $-1.048_{-0.026}^{+0.026}$ & $-1.043_{-0.029}^{+0.028}$ & $-1.047_{-0.024}^{+0.026}$ &$-3, 3$\\
$\alpha_2$ &$-1.05$ & $-1.055_{-0.027}^{+0.027}$ & $-1.064_{-0.024}^{+0.026}$ & $-1.050_{-0.027}^{+0.025}$ & $-1.053_{-0.024}^{+0.024}$  &$-3, 3$\\
$\alpha_3$ &$-1.05$ & $-1.135_{-0.181}^{+0.130}$ & $-1.046_{-0.131}^{+0.110}$ & $-1.127_{-0.157}^{+0.117}$ & $-1.103_{-0.128}^{+0.126}$ &$-3, 3$\\
$\alpha_4$ &$-1.05$ & $-1.148_{-0.447}^{+0.407}$ & $-1.146_{-0.325}^{+0.482}$ & $-0.949_{-0.441}^{+0.403}$ & $-1.111_{-0.418}^{+0.344}$ &$-3, 3$\\
$\beta_1$ &$\dots$ & $\pp3.204_{-1.933}^{+1.863}$ & $\pp2.751_{-1.461}^{+1.877}$ & $\pp2.954_{-1.665}^{+1.906}$ & $\pp2.897_{-1.188}^{+1.721}$ &$-1, 10$\\
$\beta_2$ &$\dots$ & $\pp2.983_{-1.317}^{+1.191}$ & $\pp2.503_{-0.944}^{+1.256}$ & $\pp2.715_{-1.154}^{+1.239}$ & $\pp2.569_{-0.876}^{+1.273}$ &$-1, 10$\\
$\beta_3$ &$\dots$ & $\pp0.890_{-0.253}^{+0.253}$ & $\pp0.955_{-0.220}^{+0.218}$ & $\pp1.171_{-0.245}^{+0.228}$ & $\pp1.061_{-0.210}^{+0.202}$ &$-1, 10$\\
$\beta_4$ &$\dots$ & $\pp0.518_{-0.101}^{+0.101}$ & $\pp0.380_{-0.093}^{+0.086}$ & $\pp0.377_{-0.094}^{+0.092}$ & $\pp0.367_{-0.079}^{+0.089}$ &$-1, 10$\\
$\beta_5$ &$\dots$ & $\pp0.159_{-0.063}^{+0.070}$ & $\pp0.141_{-0.052}^{+0.065}$ & $\pp0.171_{-0.060}^{+0.064}$ & $\pp0.150_{-0.055}^{+0.050}$ &$-1, 10$\\
$\beta_6$ &$\dots$ & $\pp0.060_{-0.077}^{+0.066}$ & $\pp0.041_{-0.068}^{+0.062}$ & $\pp0.048_{-0.066}^{+0.064}$ & $\pp0.046_{-0.063}^{+0.057}$ &$-1, 10$\\
$\beta_7$ &$\dots$ & $\pp0.018_{-0.048}^{+0.044}$ & $\pp0.011_{-0.049}^{+0.043}$ & $\pp0.009_{-0.025}^{+0.046}$ & $\pp0.017_{-0.030}^{+0.032}$ &$-1, 10$\\
$\logten(A_{\rm IHL_1}) $ &$\dots$ & $-1.267_{-1.495}^{+1.211}$ & $-1.208_{-1.211}^{+1.267}$ & $-1.401_{-1.428}^{+0.964}$ & $-1.407_{-1.264}^{+1.192}$ &$-10, 0$\\
$\logten(A_{\rm IHL_2}) $ &$\dots$ & $-1.429_{-0.393}^{+0.326}$ & $-1.461_{-0.306}^{+0.273}$ & $-1.392_{-0.311}^{+0.363}$ & $-1.403_{-0.258}^{+0.290}$ &$-10, 0$\\
$\logten(A_{\rm IHL_3}) $ &$\dots$ & $-1.514_{-0.089}^{+0.077}$ & $-1.562_{-0.072}^{+0.068}$ & $-1.632_{-0.077}^{+0.070}$ & $-1.623_{-0.074}^{+0.067}$ &$-10, 0$\\
$\logten(A_{\rm IHL_4}) $ &$\dots$& $-1.394_{-0.091}^{+0.086}$ & $-1.277_{-0.068}^{+0.074}$ & $-1.274_{-0.070}^{+0.079}$ & $-1.254_{-0.077}^{+0.066}$ &$-10, 0$\\
$\logten(A_{\rm IHL_5}) $ &$\dots$& $-1.026_{-0.099}^{+0.096}$ & $-0.990_{-0.088}^{+0.074}$ & $-1.038_{-0.096}^{+0.089}$ & $-1.014_{-0.070}^{+0.079}$ &$-10, 0$\\
$\logten(A_{\rm IHL_6}) $ &$\dots$& $-0.857_{-0.144}^{+0.148}$ & $-0.838_{-0.129}^{+0.130}$ & $-0.844_{-0.136}^{+0.137}$ & $-0.833_{-0.121}^{+0.124}$ &$-10, 0$\\
$\logten(A_{\rm IHL_7}) $ &$\dots$& $-0.745_{-0.164}^{+0.162}$ & $-0.707_{-0.138}^{+0.148}$ & $-0.711_{-0.154}^{+0.140}$ & $-0.733_{-0.128}^{+0.126}$ &$-10, 0$\\
\hline
\multicolumn{7}{c}{IGL}\\
$A^{\phi}_1$ &$1$ & $\pp1.006_{-0.010}^{+0.010}$ & $\pp1.005_{-0.009}^{+0.008}$ & $\pp1.009_{-0.009}^{+0.008}$ & $\pp1.005_{-0.008}^{+0.008}$ &$0, 2$\\
$A^{\phi}_2$ &$1$ & $\pp0.990_{-0.013}^{+0.012}$ & $\pp0.991_{-0.012}^{+0.011}$ & $\pp0.991_{-0.012}^{+0.011}$ & $\pp0.990_{-0.011}^{+0.012}$ &$0, 2$\\
$A^{\phi}_3$ &$1$ & $\pp1.011_{-0.019}^{+0.019}$ & $\pp1.013_{-0.017}^{+0.017}$ & $\pp1.008_{-0.019}^{+0.017}$ & $\pp1.007_{-0.017}^{+0.017}$ &$0, 2$\\
$A^{\phi}_4$ &$1$ & $\pp1.008_{-0.023}^{+0.021}$ & $\pp1.011_{-0.021}^{+0.023}$ & $\pp1.007_{-0.022}^{+0.022}$ & $\pp1.005_{-0.021}^{+0.022}$ &$0, 2$\\
$A^{\phi}_5$ &$1$ & $\pp1.004_{-0.024}^{+0.023}$ & $\pp1.008_{-0.022}^{+0.022}$ & $\pp1.002_{-0.023}^{+0.021}$ & $\pp0.999_{-0.021}^{+0.022}$ &$0, 2$\\
$A^{\phi}_6$ &$1$ & $\pp1.009_{-0.022}^{+0.023}$ & $\pp1.011_{-0.021}^{+0.021}$ & $\pp1.005_{-0.020}^{+0.022}$ & $\pp1.003_{-0.020}^{+0.021}$ &$0, 2$\\
$A^{\phi}_7$ &$1$ & $\pp1.003_{-0.022}^{+0.022}$ & $\pp1.007_{-0.019}^{+0.019}$ & $\pp0.999_{-0.020}^{+0.020}$ & $\pp0.997_{-0.019}^{+0.020}$ &$0, 2$\\
$A^{\phi}_8$ &$1$ & $\pp1.006_{-0.018}^{+0.019}$ & $\pp1.009_{-0.017}^{+0.018}$ & $\pp1.002_{-0.017}^{+0.019}$ & $\pp1.000_{-0.017}^{+0.016}$ &$0, 2$\\
$A^{\phi}_9$ &$1$ & $\pp1.012_{-0.013}^{+0.015}$ & $\pp1.015_{-0.013}^{+0.015}$ & $\pp1.010_{-0.013}^{+0.014}$ & $\pp1.009_{-0.013}^{+0.013}$ &$0, 2$\\
$A^{\phi}_{10}$ &$1$ & $\pp1.004_{-0.011}^{+0.011}$ & $\pp1.005_{-0.011}^{+0.010}$ & $\pp1.001_{-0.011}^{+0.011}$ & $\pp1.000_{-0.010}^{+0.010}$ &$0, 2$\\
\hline
\multicolumn{7}{c}{EoR}\\
$f_{*_1}$ &$\dots$ &$\pp0.023^{+0.015}_{-0.014}$ &$\pp0.022^{+0.015}_{-0.011}$ &$\pp0.024^{+0.015}_{-0.012}$ &$\pp0.022^{+0.013}_{-0.010}$ &$-1, 1$\\
$f_{*_2}$ &$\dots$ &$\pp0.015^{+0.008}_{-0.007}$ &$\pp0.021^{+0.006}_{-0.006}$ &$\pp0.022^{+0.006}_{-0.007}$ &$\pp0.014^{+0.006}_{-0.005}$ &$-1, 1$\\
$f_{*_3}$ &$\dots$ &$\pp0.014^{+0.010}_{-0.017}$ &$\pp0.012^{+0.011}_{-0.012}$ &$\pp0.013^{+0.012}_{-0.014}$ &$\pp0.011^{+0.010}_{-0.009}$ &$-1, 1$\\
$f_{*_4}$ &$0.01$ &$\pp0.012^{+0.034}_{-0.045}$ &$\pp0.015^{+0.034}_{-0.037}$ &$\pp0.011^{+0.038}_{-0.046}$ &$\pp0.013^{+0.035}_{-0.039}$ &$-1, 1$\\
\hline
\multicolumn{7}{c}{DGL}\\
$A_{\rm DGL}$ &$1$ & $\pp0.992_{-0.031}^{+0.027}$ & $\pp0.994_{-0.028}^{+0.027}$ & $\pp0.991_{-0.028}^{+0.029}$ & $\pp0.998_{-0.026}^{+0.027}$ &$0, 2$\\
\hline
\multicolumn{7}{c}{Cosmic shear}\\
$m_1$ &$0$&$\dots$& $-0.001^{+0.003}_{-0.002}$ &$\dots$&$-0.001^{+0.002}_{-0.002}$ &$-1, 1^*$\\
$m_2$ &$0$&$\dots$& $\pp0.001^{+0.003}_{-0.003}$ &$\dots$&$\pp0.001^{+0.002}_{-0.003}$ &$-1, 1^*$\\
$m_3$ &$0$&$\dots$& $\pp0.000^{+0.002}_{-0.002}$ &$\dots$&$\pp0.000^{+0.002}_{-0.002}$ &$-1, 1^*$\\
$m_4$ &$0$&$\dots$& $-0.001^{+0.002}_{-0.003}$ &$\dots$&$\pp0.000^{+0.002}_{-0.002}$ &$-1, 1^*$\\
$m_5$ &$0$&$\dots$& $\pp0.001^{+0.002}_{-0.003}$ &$\dots$&$\pp0.000^{+0.002}_{-0.002}$ &$-1, 1^*$\\
$m_6$ &$0$&$\dots$& $\pp0.001^{+0.002}_{-0.002}$ &$\dots$&$-0.001^{+0.002}_{-0.002}$ &$-1, 1^*$\\
$A_{\rm IA_1}$ &$1$&$\dots$& $\pp0.986^{+0.049}_{-0.050}$&$\dots$&$\pp0.992^{+0.049}_{-0.047}$ &$0, 2$\\
$A_{\rm IA_2}$ &$1$&$\dots$& $\pp0.967^{+0.161}_{-0.165}$&$\dots$&$\pp1.014^{+0.134}_{-0.150}$ &$0, 2$\\
\hline
\multicolumn{7}{l}{\footnotesize ${}^{*}$ The multiplicative calibration parameters $m_i$ are constrained by a Gaussian prior $\mathcal{N}\left(0, 0.003^2\right)$.}
\end{longtable}

\end{appendix}

\end{document}